\documentclass[%
 reprint,
 superscriptaddress,
 amsmath,amssymb,
 aps,
]{revtex4-2}

\usepackage{graphicx}
\usepackage{dcolumn}
\usepackage{bm}
\usepackage{lineno}

\usepackage{xcolor}
\usepackage{gensymb}
\usepackage{pdfcomment}

\begin{document}

\preprint{APS/123-QED}

\title{Kinetic Inductance Parametric Converter}

\author{M. Khalifa}
\affiliation{Stewart Blusson Quantum Matter Institute, University of British Columbia, Vancouver, BC, Canada}\affiliation{Department of Electrical and Computer Engineering, University of British Columbia, Vancouver, BC, Canada.}
 \author{P. Feldmann}
\affiliation{Stewart Blusson Quantum Matter Institute, University of British Columbia, Vancouver, BC, Canada} \affiliation{Department of Physics and Astronomy, University of British Columbia, Vancouver, BC, Canada} 
\author{J. Salfi}%
\email{jsalfi@ece.ubc.ca}
\affiliation{Stewart Blusson Quantum Matter Institute, University of British Columbia, Vancouver, BC, Canada}\affiliation{Department of Electrical and Computer Engineering, University of British Columbia, Vancouver, BC, Canada.}\affiliation{Department of Physics and Astronomy, University of British Columbia, Vancouver, BC, Canada}

\date{\today}

\begin{abstract}
Parametric converters are parametric amplifiers that mix two spatially separate nondegenerate modes and are commonly used for amplifying and squeezing microwave signals in quantum computing and sensing. 
In Josephson parametric converters, the strong localized nonlinearity of the Josephson Junction limits the amplification and squeezing, as well as the dynamic range, in current devices. 
In contrast, a weak distributed nonlinearity can provide higher gain and dynamic range, when implemented as a kinetic inductance (KI) nanowire of a dirty superconductor, and has additional benefits such as resilience to magnetic field, higher-temperature operation, and simplified fabrication. Here, we propose, demonstrate, and analyze the performance of a KI parametric converter that relies on the weak distributed nonlinearity of a KI nanowire. The device utilizes three-wave mixing induced by a DC current bias. We demonstrate its operation as a nondegenerate parametric amplifier with high phase-sensitive gain, reaching two-mode amplification and deamplification of $\sim$30 dB for two resonances separated by 0.8 GHz, in excellent agreement with our theory of the device.  We observe a dynamic range of -108~dBm at 30 dB gain. Our device can significantly broaden applications of quantum-limited signal processing devices including phase-preserving amplification and two-mode squeezing. 
\end{abstract}

\maketitle

\section{\label{sec:Intro}Introduction}

Superconducting quantum optics finds many potential technological applications including continuous-variable quantum computation \cite{ofek2016extending,chapman2023high,he2023fast} and teleportation \cite{pogorzalek2019secure,fedorov2021experimental}, remote qubit entanglement\cite{silveri2016theory}, and quantum illumination \cite{barzanjeh2020microwave}. In these applications, entangled photon pairs or two-mode squeezed states over two separate microwave channels are important\cite{flurin2012generating,pogorzalek2019secure}.
This is typically accomplished using two Josephson parametric amplifiers~\cite{pogorzalek2019secure} or a Josephson parametric converter (JPC)~\cite{bergeal2010analog,abdo2013nondegenerate} in which a few discrete Josephson Junctions (JJs) provide the nonlinearity. The JPC has further applications including frequency conversion and beam splitting \cite{abdo2013full}, directional amplification \cite{sliwa2015reconfigurable,abdo2017gyrator,chien2020multiparametric}, qubit readout \cite{hatridge2013quantum,abdo2021high}, and two-qubit coupling \cite{leib2016transmon,roy2017implementation}. However, recent studies have shown fundamental limitations on the JPC's amplification and squeezing capabilities due to the higher order effects of the strong localized nonlinearity of JJs \cite{liu2017josephson,kim2023squeezing}.
Parametric amplifiers with weak distributed nonlinearity based on kinetic inductance (KI) have been shown experimentally to possess favourable features including higher gain and dynamic range \cite{parker2022degenerate}, strong single-mode squeezing \cite{vaartjes2023strong}, and operation at high magnetic fields~\cite{khalifa2023nonlinearity,xu2023magnetic, frasca2024three,splitthoff2024gate} and temperatures \cite{malnou2022performance,mohamed2023selective}. 

Here, we introduce and experimentally demonstrate a KI parametric converter (KIPC) made of the distributed weak nonlinearity provided by the KI of a superconducting nanowire. We find that the device provides high gain on two spatially separated modes and exhibits high dynamic range.

The KIPC is a four-port device, where each port selectively couples to one of the two modes, unlike previously reported two-mode kinetic inductance parametric amplifiers, where the two modes couple to the same ports \cite{wu2023junction,mohamed2023selective}.
To separate the frequencies of the resonance modes and to realize selective coupling, the design breaks the continuous symmetry of the ring by varying the KI-enhanced impedance for different sections of the ring.
For single-mode input, we achieve two-mode amplification above 30 dB with dynamic range around ${-108}$ dBm, exceeding the dynamic range of previously reported JPCs at the same gain by more than 20 dB~\cite{bergeal2010phase},
and confirm cross-correlation between the amplified quadratures of the output signal and idler.
We also observe interference fringes for two-mode input with deamplification of 27 dB and output coherent cancellation of 99.995\%.

We expect our device, which features high dynamic range and simple fabrication and promises higher temperature and magnetic-field operation, to find applications in phase-sensitive amplification and two-mode squeezing. The weak distributed nonlinearity inherent in our design could be helpful to obtain higher amplification and squeezing than in devices with localized nonlinearities in which higher harmonics may play a role\cite{liu2017josephson,kim2023squeezing}.

\section{\label{sec:Dev} Device}

\begin{figure*}
    \centering
    \includegraphics [width=0.66\linewidth] {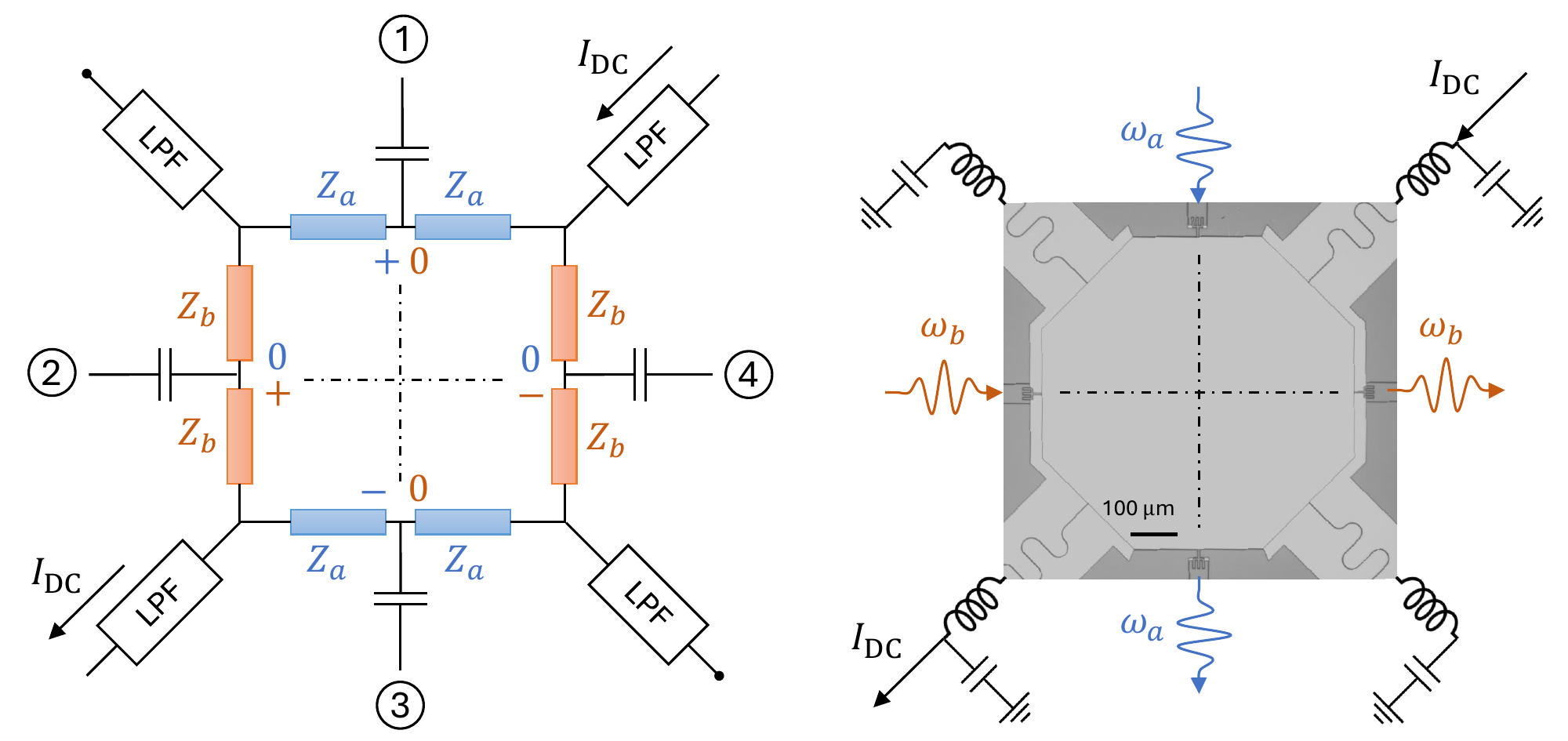}
    \includegraphics [width=0.3\linewidth] {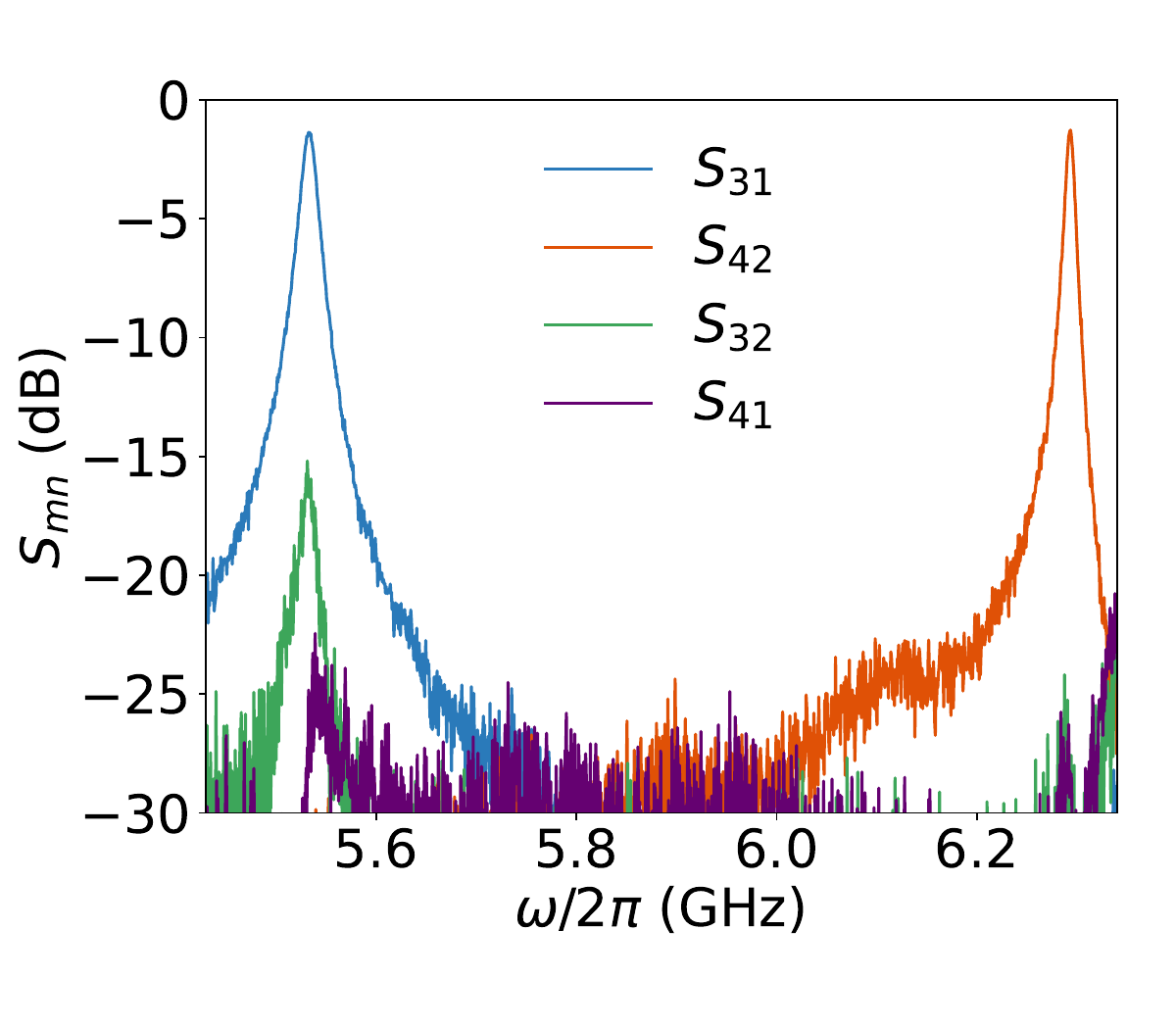}
    \put(-490,147){(a)}
    \put(-310,147){(b)}
    \put(-150,147){(c)}
    \caption{Device design and linear characterization. (a) Schematic of the KIPC. It consists of a KI nanowire ring resonator with two symmetry axes defined by sections with characteristic impedance $Z_a\neq Z_b$. The ring has four microwave ports (1 to 4) aligned with the symmetry axes for capacitive coupling, and four DC lines for current bias through low pass filters. The symmetry constraints break the degeneracy of the two orthogonal modes in the ring such that each mode selectively couples to two ports. The \{$0$\} and \{$+$,$-$\} indicate the voltage nodes and antinodes, respectively, for mode $a$ (blue) and mode $b$ (orange). 
    (b) Optical image of the fabricated device. The ring has an octagon shape. The impedance is varied by changing the distance between the nanowire and the ground plane to get $Z_a\approx940$ $\Omega$ and $Z_b\approx1320$ $\Omega$. The low-pass filters are on-chip LC filters. The DC current is applied between two of the DC lines, while the other two are kept open. The coherent pump tone can be applied via any of the microwave ports.
    (c)~Transmission spectrum from ports 1 and 2 to ports 3 and 4. Mode $a$ ($b$) shows high transmission only for $S_{31}$ ($S_{42}$), confirming the selective coupling to the ports.
    }
    \label{fig:Dev}
\end{figure*}

A parametric converter possesses two nondegenerate resonant modes that couple via a three-wave mixing (3WM) process when the device is pumped near the sum or difference of their frequencies that is typically induced by a DC bias that also tunes the mode frequencies \cite{bergeal2010analog}. Here, we focus on the amplification regime, where the pump is at the sum frequency.
For a classical pump at $\omega_p$, the Hamiltonian describing this regime in a frame rotating at $\omega_p/2$ is
\begin{equation}\label{eq:hamiltonian}
    H = \hbar \Delta_a a^\dagger a + \hbar \Delta_b b^\dagger b + \frac{\hbar}{2} \left( \xi a^\dagger b^\dagger + \xi^* a b \right),
\end{equation}
where $a$ and $b$ are the annihilation operators of the two modes with resonance frequencies $\omega_a$ and $\omega_b$, respectively, $\Delta_a=\omega_a-\omega_p/2$, $\Delta_b=\omega_b-\omega_p/2$, and $\xi$ is the 3WM rate representing the strength of the nonlinearity.
The four-wave mixing (4WM) is ignored in Eq.~\eqref{eq:hamiltonian} due to the small Kerr nonlinearity of kinetic inductance nanowires~\cite{parker2022degenerate,khalifa2023nonlinearity}. 

A schematic for our device, which features two non-degenerate resonance modes $a$ and $b$ that couple to different ports, is shown in Fig. \ref{fig:Dev}(a). The device consists of a superconducting nanowire ring, capacitively coupled to four ports at its four sides. 
The characteristic impedance of the nanowire changes between two values; $Z_a$ near ports 1 and 3, and $Z_b$ near ports 2 and 4. The impedance variation along the ring enables the definition of two modes which couple predominantly to different ports of the device. As illustrated in Fig.~\ref{fig:Dev}(a), the ring exhibits two mirror symmetries, vertical and horizontal, rendering the voltage distribution of mode $a$ to be symmetric along the vertical axis and antisymmetric along the horizontal axis, and vice versa for mode $b$. The alignment of ports 1 and 3 along the vertical axis, where mode $a$ has maximum voltage and mode $b$ has zero voltage, gives maximum coupling of these ports to mode $a$, and maximum isolation from mode $b$. Similarly, ports 2 and 4 are maximally coupled to mode $b$ and maximally isolated from mode $a$. 
The 3WM nonlinearity responsible for the parametric amplification is generated by the DC current applied to the ring via low-pass filters which reflect photons in resonant modes $a$ and $b$.

The variation of the impedance over the ring separates the two modes energetically. The mode with maximum voltage at the points of high impedance has higher resonance frequency than the other mode. The resonance frequencies in the linear regime can be found using the telegrapher equation subject to symmetry-defined boundary conditions. 
The resonance frequencies of the ideal circuit in Fig.~\ref{fig:Dev}(a) obey (see Appendix \ref{app:resF})
\begin{subequations}\label{eqn:res}
\begin{equation}
    \tan\left( \frac{\omega_a}{v_a} \frac{l}{8} \right) \tan\left( \frac{\omega_a}{v_b} \frac{l}{8} \right) = \frac{Z_a}{Z_b},
\end{equation}
\begin{equation}
    \tan\left( \frac{\omega_b}{v_a} \frac{l}{8} \right) \tan\left( \frac{\omega_b}{v_b} \frac{l}{8} \right) = \frac{Z_b}{Z_a},
\end{equation}
\end{subequations}
where $l$ is the total length of the ring, while $v_a$ and $Z_a$ ($v_b$ and $Z_b$) are the phase velocity and characteristic impedance of the blue (orange) sections in Fig.~\ref{fig:Dev}(a). 
The different-impedance condition ($Z_a \neq Z_b$) is sufficient for lifting the degeneracy of the two modes. The impedance can be changed by varying a geometrical parameter in the cross-section of the nanowire~\cite{eom2012wideband}. 

We implemented the device as an octagonal ring, shown in Fig. \ref{fig:Dev}(b), made of NbTiN nanowire of thickness 10~nm, width 0.4~$\mu$m, and total length 2.2~mm. The sheet kinetic inductance of the film is $\approx40$ pH/$\square$. 
The impedance variation is realized by varying the gap between the nanowire and the ground plane \cite{goppl2008coplanar,samkharadze2016high}. At the upper and lower sides, the gap is 1 $\mu$m giving a predicted impedance $Z_a\approx940$ $\Omega$, while at the left and right sides, the gap is 25 $\mu$m giving an estimated impedance $Z_b\approx1320$ $\Omega$~\cite{samkharadze2016high}.
The low-pass filters at the four corners are on-chip superconducting LC filters implemented as kinetic inductors and inter-digital capacitors. The filters reflect photons above 0.7 GHz back into the ring and pass low frequencies.

\begin{figure*}
    \centering
    \includegraphics [width=0.33\linewidth] {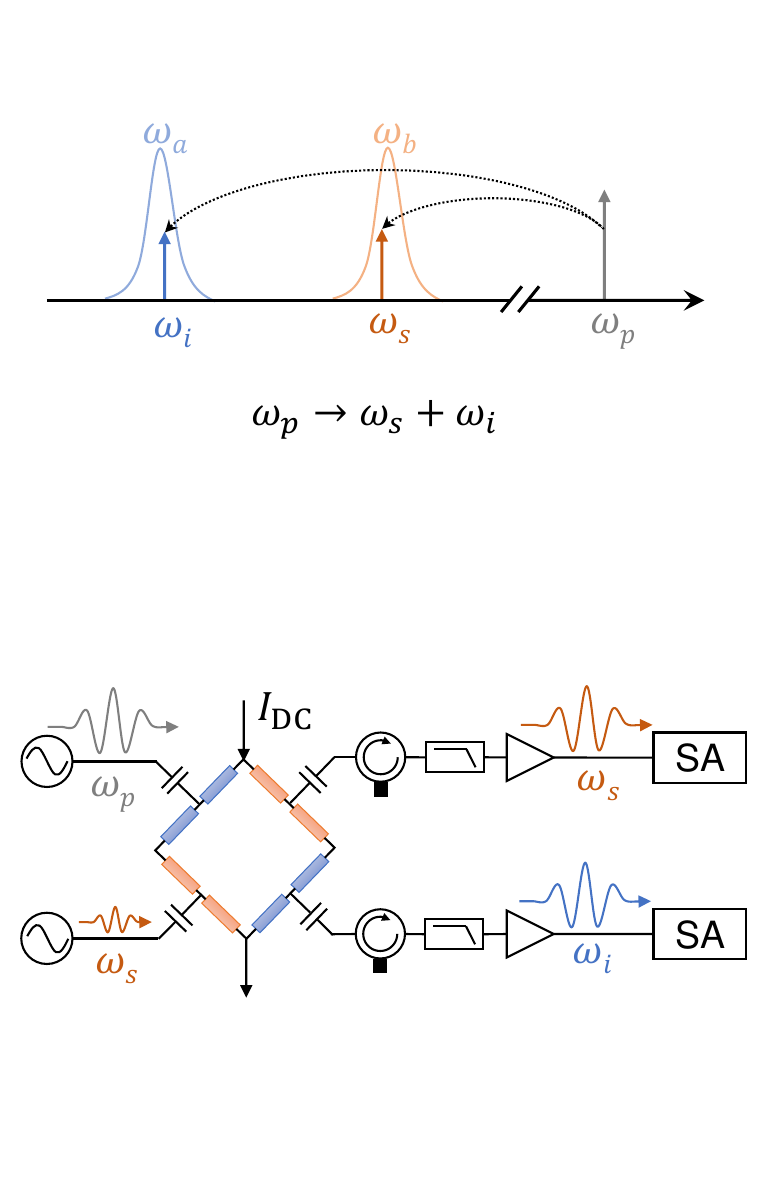}
    \includegraphics [width=0.66\linewidth] {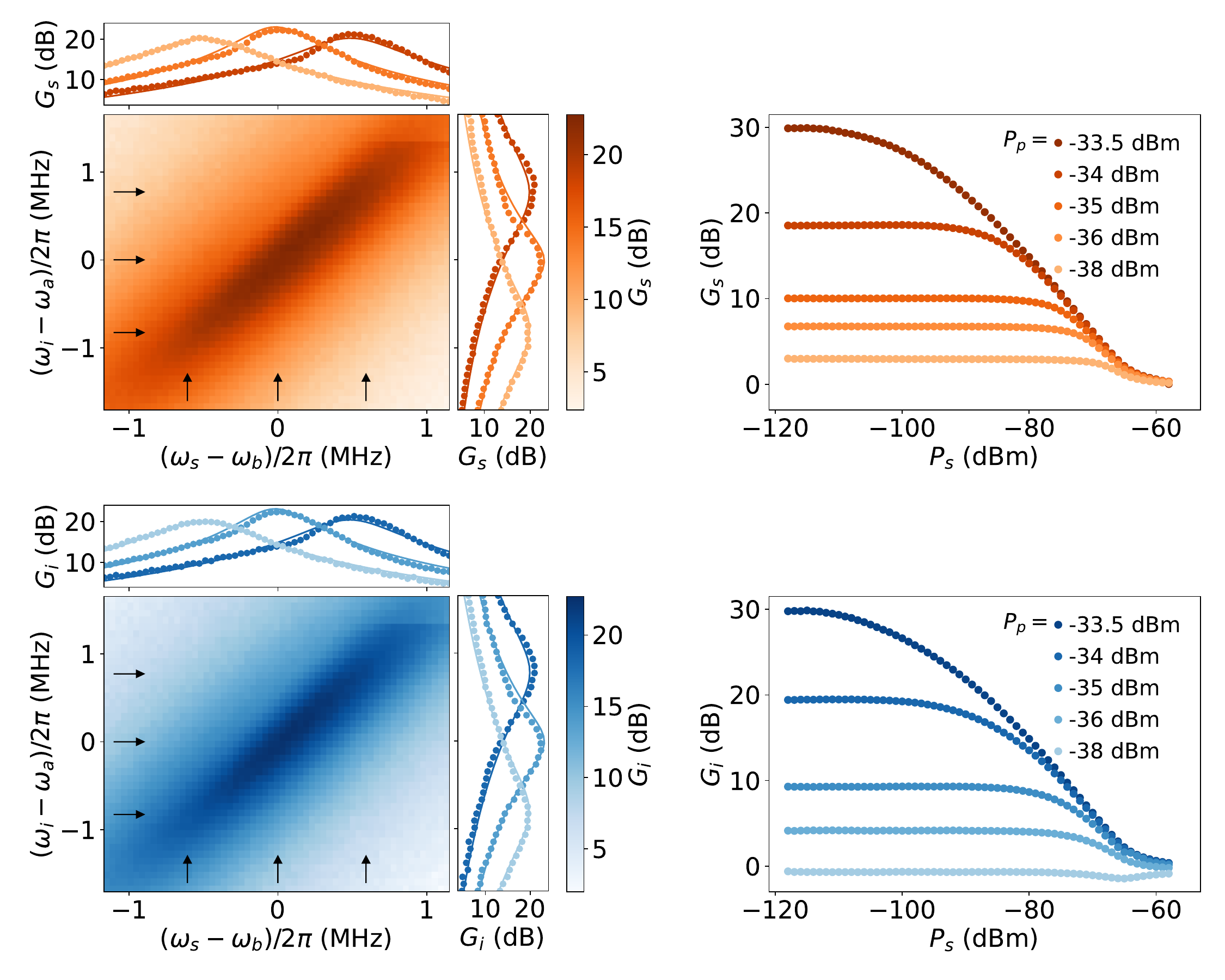}
    \put(-507,260){(a)}
    \put(-507,125){(b)}
    \put(-342,260){(c)}
    \put(-342,125){(d)}
    \put(-148,260){(e)}
    \put(-148,125){(f)}
    \caption{
    Two-mode amplification. (a) Frequency configuration for 3WM in which a pump photon is converted into signal and idler photons.
    (b) Experimental setup for amplification measurement. Signal and pump tones at frequencies $\omega_s$ and $\omega_p$ are sent to the inputs of the KIPC and signal and idler tones at $\omega_s$ and $\omega_i=\omega_p-\omega_s$ are measured at the outputs. Low-pass filters with a cutoff frequency of 8~GHz are added at the outputs to filter out the pump tone. 
    (c),(d)~Measured signal and idler gains at $\omega_s$ and $\omega_i$, respectively, versus signal detuning from mode $b$ ($\omega_s-\omega_b$) and idler detuning from mode $a$ ($\omega_i-\omega_a$) at a pump power of -34 dBm.
    Top and right panels compare experiment (points) and theory (lines, Eq.~\eqref{eqn:gain}) for slices of the 2D gain maps along the horizontal and vertical arrows, respectively.
    (e),(f)~Measured signal and idler gains, respectively, versus signal power $P_s$ at different pump powers. 
    }
    \label{fig:Amp}
\end{figure*}
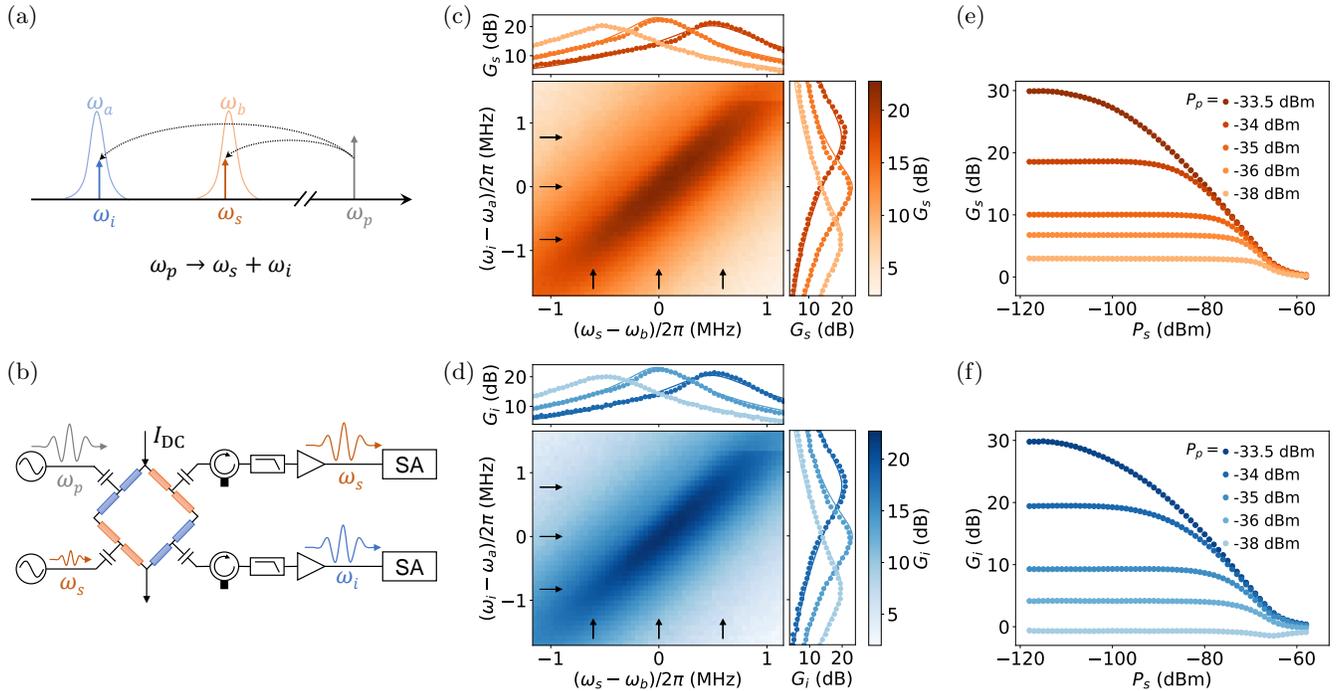

The resonance frequencies of the two modes are experimentally obtained from the transmission spectrum $S_{31}$ ($\omega_a=5.5$ GHz) and $S_{42}$ ($\omega_b=6.3$ GHz), as presented in Fig. \ref{fig:Dev}(c). The resonance frequencies predicted by Eq.~(\ref{eqn:res}) are around $4.5$~GHz and $5.5$~GHz, which are close to the measured values. The mismatch between experiment and theory probably occurs because the ground areas surrounding the filters are removed, reducing the capacitance and increasing the resonant frequency of the actual device with respect to the model.
By comparing the four transmission combinations in Fig.~\ref{fig:Dev}(c), we find that mode $a$ has 15 and 25 dB isolation from ports 2 and 4, respectively, while mode $b$ is isolated by more than 25 dB from ports 1 and 3. The residual coupling of mode $a$ to the input port for mode $b$ is likely due to deviations of the physical realization [Fig.~\ref{fig:Dev}(b)] from the idealized circuit [Fig.~\ref{fig:Dev}(a)]. Fabrication imperfections, such as variations in the sheet kinetic inductance, introduce symmetry-breaking perturbations, which might cause small shifts in the locations of voltage nodes and lead to imperfect isolation. 
The resonance frequencies are shifted by the DC current that provides the 3WM in the device (see Appendix~\ref{app:resIDC}). 
At DC current above 370 $\mu$A the device turns normal. For the nonlinear operation in the rest of the paper, we use $I_{\rm DC} = 320$ $\mu$A, leaving a margin for the microwave current from the pump tone.

\section{Results and Discussion}\label{sec:results}
\subsection{Amplification}
We operate the device as a nondegenerate parametric amplifier by sending a signal tone to port 2 at angular frequency $\omega_s \approx \omega_b$ and a pump tone to port 1 at $\omega_p \approx \omega_a+\omega_b$, and measuring the output amplified signal at $\omega_s$ and output idler at $\omega_i=\omega_p-\omega_s \approx \omega_a$. In this regime, the signal is amplified by downconverting pump photons into signal and idler photons, as illustrated in Fig.~\ref{fig:Amp}(a). A schematic of the experimental setup is shown in Fig.~\ref{fig:Amp}(b). The signal gain is extracted from the measured output powers as the ratio between the output power when pump is on, and the output power when pump is off for a signal at resonance, which represents the input power at resonance. The signal gain is extracted as the ratio between the signal output and input powers. To estimate the latter, we switch off the pump and measure the signal output power for resonant signal input. Similarly, the idler gain is referred to the input signal.

For a fixed pump power of -34 dBm, we plot the measured signal and idler gains $G_s$ and $G_i$ in Figs.~\ref{fig:Amp}(c) and (d), respectively.  Data was taken for different detunings of the signal (idler) from resonant mode $b$ ($a$) by varying $\omega_s$ and $\omega_p$. The signal gain [Fig.~\ref{fig:Amp}(c)] peaks around the two-mode resonance point, and reduces gradually far from resonance with tilted elliptic contours. A similar pattern is observed for the idler gain [Fig.~\ref{fig:Amp}(d)] with the same value and location for the maximum gain point. For measurements of sufficiently long duration, e.g., in Figs.~\ref{fig:Amp}(c) and (d), we observe jumps in the gain of the device. These are more frequent at higher pump powers or without filtering the current bias. All measurements presented in this paper were conducted with a 10 kHz low-pass filter at room temperature on the current bias line. 

A theoretical analysis of the device using input-output theory (see Appendix \ref{app:inout}) predicts signal and idler amplitude gains
\begin{subequations}\label{eqn:gain}
\begin{equation}
    g_s(\delta) = \frac{\kappa_b}{D_s(\delta)} \left[ i(\delta + \Delta_a) - \kappa_a \right],
\end{equation}
\begin{equation}
    g_i(-\delta) = i\frac{\sqrt{\kappa_a\kappa_b}}{D_i(-\delta)} \frac{\xi}{2},
\end{equation}
\end{subequations}
where $\delta=\omega_s-\omega_p/2$, $\kappa_a$ ($\kappa_b$) is the coupling loss rate of mode $a$ ($b$) to ports 1 and 3 (2 and 4), and the denominator is given by $D_{s/i}(\delta)=\left[ i(\delta \pm \Delta_a) - \kappa_a \right] \left[ i(\delta \mp \Delta_b) - \kappa_b \right] - |\xi|^2/4$. The expressions in Eq.~\eqref{eqn:gain} are related to the measured power gains by $G_{s/i}=|g_{s/i}|^2$. 
We fit the measured signal [idler] gain in Fig.~\ref{fig:Amp}(c) [Fig.~\ref{fig:Amp}(d)] to $G_s$ [$G_i$], to extract  $|\xi|=7.408(32)$ MHz [7.729(31) MHz], $\kappa_a=4.597(26)$ MHz [4.829(25) MHz], and $\kappa_b=3.210(18)$ MHz [3.315(18) MHz].

In addition to the similar extracted couplings with small fitting uncertainties, the top and side panels in Figs.~\ref{fig:Amp}(c) and (d) illustrate the good agreement between theory and experiment. This demonstrates the ability of our device to perform nondegenerate 3WM and to generate the output into two separate ports, which is the primary function of a parametric converter~\cite{bergeal2010phase,abdo2013nondegenerate}.

We characterize the gain and dynamic range of the device at different pump powers. Figures~\ref{fig:Amp}(e) and (f) show the measured signal and idler gains, respectively, versus signal power, $P_s$, at different pump powers. 
At high gain, the signal and idler have similar gain levels, while at low gain ($<10$ dB), where the input signal represents a significant part of the output, the idler gain is smaller than the signal gain. The 1-dB compression points of the signal and idler gains are similar, and they are 22-32 dB larger than the typical values reported for the JPC \cite{bergeal2010phase, abdo2011josephson, roch2012widely}. For example, for a signal gain of 30 dB (20 dB), our KIPC exhibits 1-dB compression at $P_s \sim -108$ dBm ($\sim -93$ dBm), while previously reported JPCs have 1-dB compression points around $-130$ dBm \cite{bergeal2010phase} ($-125$ dBm \cite{abdo2011josephson}).

\begin{figure}
    \centering
    \includegraphics [width=0.8\linewidth]{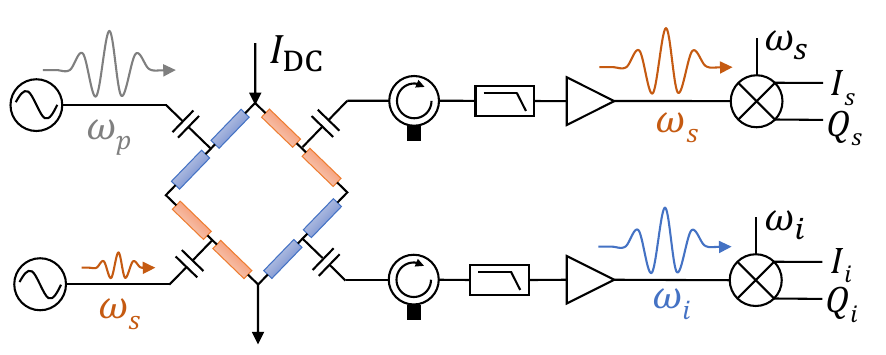}\raisebox{24mm}{\makebox[0pt][l]{r\hspace{-72.2mm}(a)}}
    \includegraphics [width=0.9\linewidth]{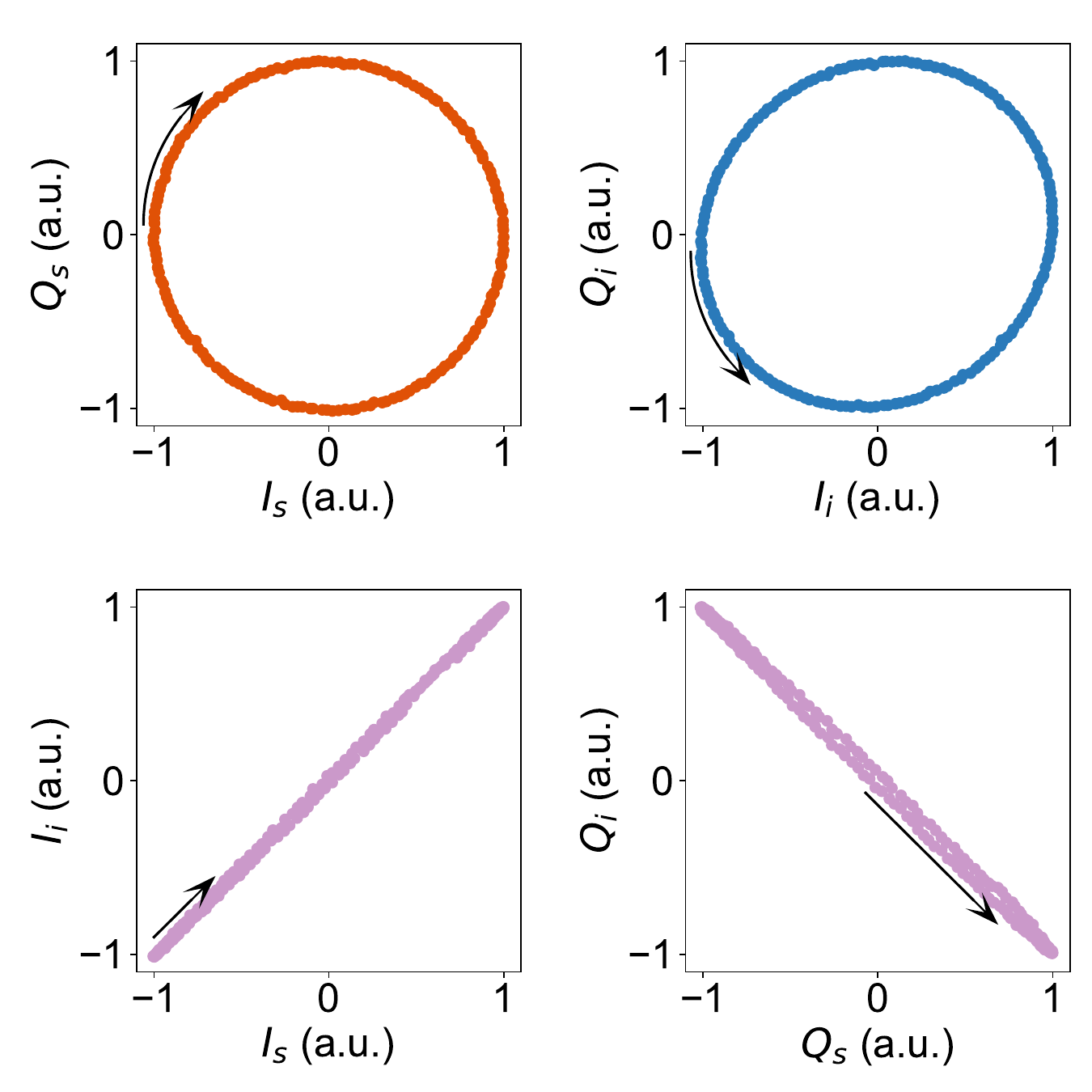}
    \put(-218,215){(b)}
    \put(-106,215){(c)}
    \put(-218,105){(d)}
    \put(-106,105){(e)}
    \caption{Phase correlation. (a) Basic experimental setup for measuring the output quadratures. (b),(c)~Scatter plots for the measured quadratures of the output signal and idler, respectively, at varying input signal phase $\phi_s \in[-\pi,\pi]$. The results are normalized, and the phases of the outputs are shifted so that $(I,Q)=(-1,0)$ at $\phi_s=-\pi$. (d) Scatter plot for the $I$ components of the signal and idler, showing positive correlation. (e) Scatter plot for the $Q$ components of the signal and idler, showing anticorrelation. All arrows start at $\phi_s=-\pi$ and show the direction of increasing $\phi_s$.}
    \label{fig:Phase}
\end{figure}

\subsection{Phase Correlation}
Next, we investigate the correlation between the phases of the output signal and idler by varying the input signal phase while keeping the pump phase fixed.
The phase correlation is determined by measuring the four quadratures of the two outputs as shown in Fig. \ref{fig:Phase}(a).
By demonstrating the output phase correlation for a coherent input signal, we verify the phase coherence of the 3WM process in our device. Accordingly, we expect to generate phase-correlated pairs when the device is operated on the vacuum state.

Scatter plots for different combinations of the four output quadratures are presented in Fig. \ref{fig:Phase}. 
The input signal phase is varied from $-\pi$ to $\pi$, which corresponds to a full circle in the input phase space. 
Figure \ref{fig:Phase}(b) shows the full circle on the signal output $I_s$-$Q_s$ phase diagram, verifying the phase-preserving nature of the KIPC, similar to the JPC~\cite{bergeal2010phase}. 
The output idler components $I_i$-$Q_i$ also exhibit a full circle [Fig. \ref{fig:Phase}(c)], indicating the dependence of the idler phase on the input signal phase.
Figures~\ref{fig:Phase}(e) and (d) show that $I_s$ and $I_i$ are positively correlated, while $Q_s$ and $Q_i$ are anticorrelated, which agrees with theory (Appendix \ref{app:inout}) and matches the cross-correlation properties of the JPC \cite{flurin2012generating}. 

\subsection{Deamplification}

\begin{figure*}
    \centering
    \includegraphics [width=0.22\linewidth] {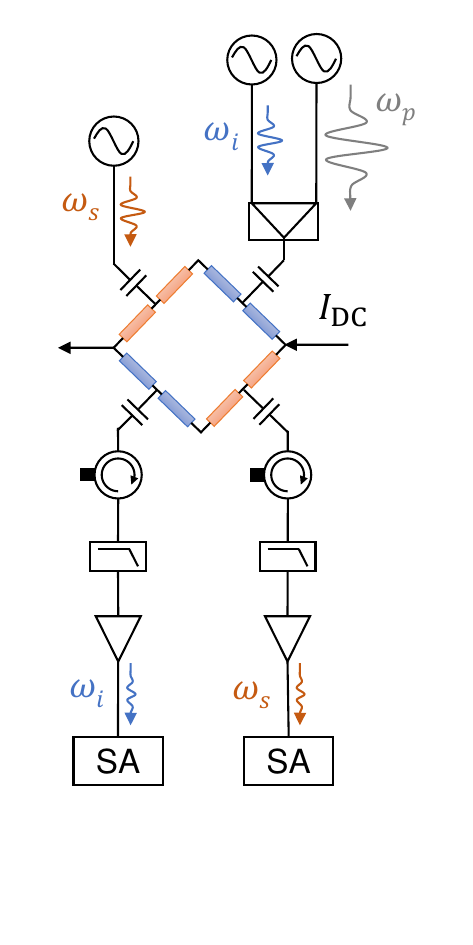}
    \includegraphics [width=0.77\linewidth] {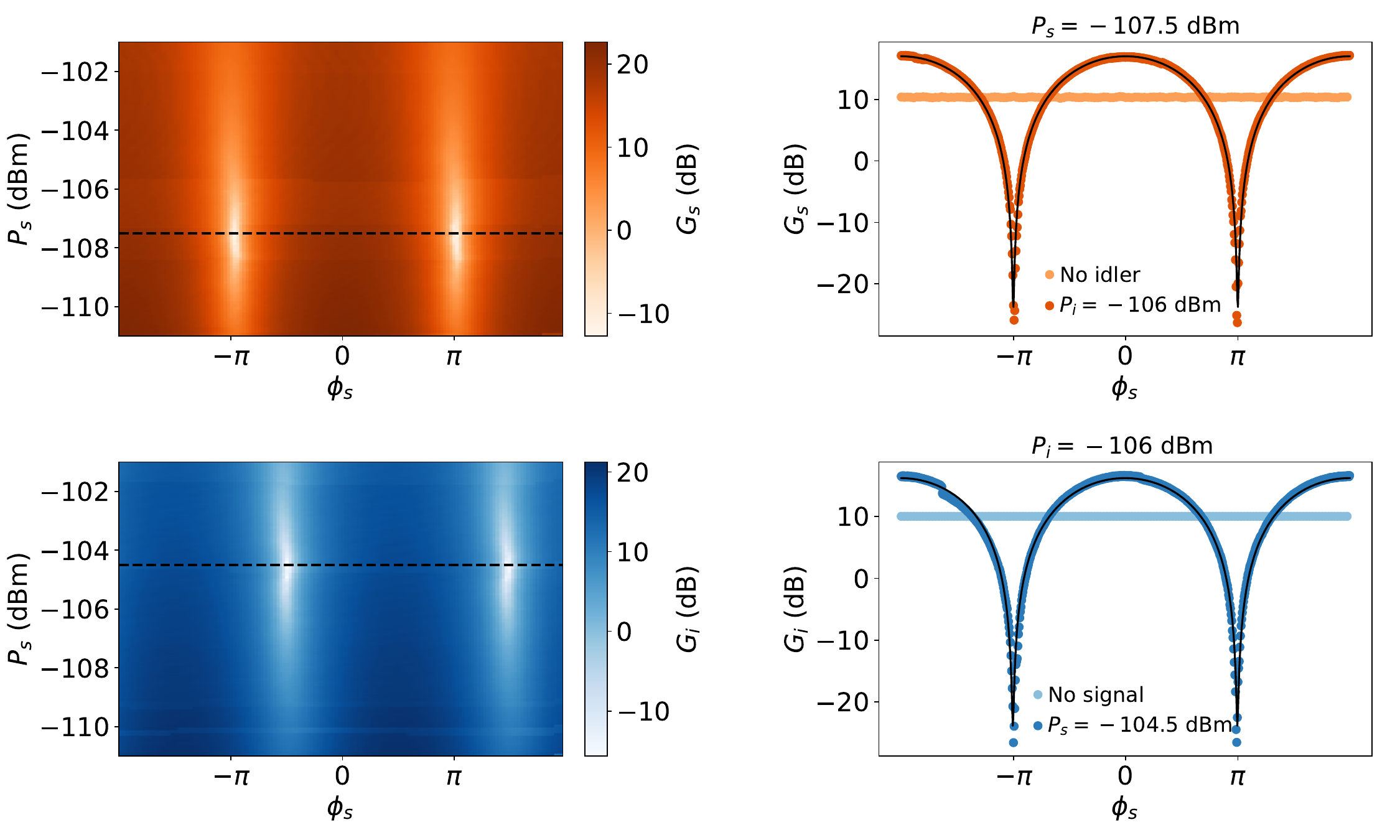}
    \put(-492,227){(a)}
    \put(-394,227){(b)}
    \put(-394,110){(c)}
    \put(-175,227){(d)}
    \put(-175,110){(e)}
    \caption{Two-mode interference and deamplification. 
    (a) Basic experimental setup for interference and deamplification measurement. Pump and idler tones are combined and sent to the device through port 1, while the signal tone is sent through port~2. Low-pass filters remove the pump from the outputs. 
    (b),(c) Signal and idler gains relative to the input signal, respectively, versus signal phase and power. The input idler power is fixed at $P_i=-106$ dBm. The dashed lines highlight the signal power level corresponding to maximum deamplification. 
    (d),(e) Signal and idler gain versus signal phase at $P_i=-106$ dBm and optimal $P_s$ [dashed lines in (b),(c)]. The measurement is taken at high phase resolution (1$\degree$) and low measurement bandwidth (100 Hz). The solid lines are obtained from theory [Eq.~\eqref{eqn:gain}] using $\kappa_a=4.55$ MHz (4.96 MHz), $\kappa_b=3.57$ MHz (3.41 MHz), $\xi=7.00$ MHz (7.45 MHz), and $P_i/P_s=1.28$ ($0.84$) for the signal (idler). The fitted couplings resemble those obtained from Fig.~\ref{fig:Amp}, and the extracted ratios $P_i/P_s$ are close to the experimentally estimated values of 1.41 and 0.71. The horizontal lines at about 10 dB indicate the signal gain (idler gain with respect to idler input) at no idler (signal) input. The pump power for all data is -35 dBm. We expect that the discontinuity in gain in Fig.~\ref{fig:Deamp}(e) has a common cause with the discontinuities observed in Fig.~\ref{fig:Amp}, discussed in Sec.~\ref{sec:results}A. }
    \label{fig:Deamp}
\end{figure*}

Parametric converters can be operated in the coherent attenuation regime, where incoming signal and idler tones are deamplified at a certain relative phase \cite{schackert2013three}. For parametric converters with hybrid couplers, as the JPC is usually used \cite{flurin2012generating,abdo2013nondegenerate,sliwa2015reconfigurable}, the deamplification results from upconverting incident signal and idler photons into pump photons. In that case, the device has effectively three ports that couple selectively to the signal, idler, and pump, where the input and output of each tone share the same port (reflection regime), and the deamplification level is limited by the single-input gain $G_0$ (maximum deamplification is $\sim4G_0$~\cite{schackert2013three}). Although our device can be utilized in a similar way by adding a hybrid coupler, we focus here on investigating the deamplification in the transmission regime without the hybrid coupler. In this case, the deamplification at the output port can result from partial upconversion of incident photons or from destructive interference at the output port, depending on the input signal and idler powers and phases. In the latter case, the output at any given port can in principle vanish at arbitrary gain (see Appendix \ref{app:inout}). Energy is conserved by total unamplified reflection of the corresponding input without net generation of pump photons. This feature can be used for routing and splitting microwave signals on chip \cite{sliwa2015reconfigurable}.

We test the deamplification of the KIPC by interfering two tones at the signal and idler frequencies.
Figure \ref{fig:Deamp}(a) shows the setup for this experiment. Here, we insert coherent input tones at both $\omega_s$ and $\omega_i$, in addition to the pump tone at $\omega_p=\omega_s+\omega_i$. We vary the signal phase, while keeping the pump and the input idler phases fixed. We also sweep the signal power to reach maximum deamplification at fixed idler and pump powers. 
Figures \ref{fig:Deamp}(b) and (c) show the measured signal and idler gains (relative to the input signal power), respectively, versus signal phase and power. We observe, for both $G_s$ and $G_i$, interference fringes and a power dependence of the deamplification. 

To quantify the deamplification more precisely, we set the signal power to the value that minimizes $G_s$ or $G_i$, and measure the respective gain with higher phase resolution and lower measurement bandwidth. For both output ports, we reach a deamplification of 27 dB, as shown in Figs.~\ref{fig:Deamp}(d) and (e). 
By fitting the data in Figs.~\ref{fig:Deamp}(d) and (e), we find similar couplings to those obtained from Fig.~\ref{fig:Amp}.
When any of the input tones is turned off, the interference pattern disappears and we get a phase-independent gain of 10 dB with respect to the remaining input.
Figures~\ref{fig:Deamp}(d) and~(e) also show that, at constructive interference both $G_s$ and $G_i$ are 6 dB above the single-input level, indicating a twofold increase in the voltage gain (fourfold increase in the power gain) at these points. This gives a ratio of $-43$ dB between the outputs at destructive and constructive interference, which corresponds to a coherent cancellation of $99.995\%$ for the output signal and idler tones \cite{schackert2013three,abdo2013full}.
We note that the measured deamplification level  of 27 dB is higher than the expected value for upconversion, given the single-input gain of 10 dB. This is consistent with input-output theory predicting that the deamplification shown in Figs. \ref{fig:Deamp}(d) and (e) results from destructive interference rather than upconversion.

\section{Conclusion}
We have designed, fabricated, and tested a novel KI parametric converter, providing the functionality of the JPC, but differing from the JPC in its use of distributed weak KI nonlinearity, instead of the nonlinearity of a few discrete JJs. The symmetry properties of our device enable the spatial and temporal separation of its two modes, while maintaining simple design and fabrication. 
We have demonstrated the basic operation of the device as a nondegenerate parametric amplifier, and measured signal and idler gain above $30$ dB and dynamic range of around $-93$ dBm at $20$ dB gain. 
In addition, we have observed cross-correlation between the quadratures of signal and idler, suggesting the possibility of generating entangled microwave beams and two-mode squeezing. For coherent signal and idler inputs, we have obtained a strong 27 dB deamplification due to destructive interference, which can be used to route and split microwave signals. The device properties are in good agreement with input-output theory model we developed for the device.

These results set the KIPC as a potential alternative for the widely used JPC, generally for the simplicity of its design and fabrication, and particularly in the applications requiring high input power, high temperature or magnetic field. These advantages result from its high dynamic range, and the inherently anticipated operation above 1 K and resilience to magnetic field. 
This work motivates further studies on the noise properties and two-mode squeezing capabilities of the KIPC, relative to the JPC and other Josephson-based approaches for two-mode squeezing and entangled photon-pair generation.

\begin{acknowledgments}
The authors thank Shabir Barzanjeh and Abdul Mohamed for helpful discussions and comments. This work was undertaken with support from the Stewart Blusson Quantum Matter Institute (SBQMI) and the Canada First Research Excellence Fund, Quantum Materials and Future Technologies Program. We acknowledge financial support from MITACs Accelerate and Syniad Innovations, the National Science and Engineering Research Council of Canada (NSERC) Discovery Grants program, and the Canadian Foundation for Innovation (CFI) John Edwards Leaders Foundation (JELF). MK acknowledges financial support from the SBQMI QuEST fellowship program. The authors acknowledge CMC Microsystems for the provision of computer aided design tools that were essential to obtain the results presented here. This research was supported in part through computational resources provided by Advanced Research Computing at the University of British Columbia.
\end{acknowledgments}

\appendix

\section{Resonance Frequencies of the KIPC}\label{app:resF}
We theoretically calculate the resonance frequencies of the two fundamental modes for the KIPC ideal circuit in Fig.~\ref{fig:Dev}(a), which consists of four transmission line sections (1 to 4) of equal length, $l/4$. Each section $j$ has uniform inductance and capacitance per unit length, $L_j$ and $C_j$, where $L_j=L_a$ and $C_j=C_a$ for $j=1,3$, while $L_j=L_b$ and $C_j=C_b$ for $j=2,4$. We consider the telegrapher model for each section of the ring separately, and apply boundary conditions at the intersection point between two adjacent sections by requiring continuous voltage and current. Due to the device symmetry, we need to apply the boundary conditions at only one intersection point. The dependence of the inductance on the microwave current is ignored. Also, the coupling capacitors and the low-pass filters are ignored.

The temporal charge distribution of mode $m \in \{a,b\}$ along section $j$ can be described by the charge telegrapher equation as \cite{parker2022degenerate}
\begin{equation}
    \frac{\partial^2 Q_m(x_j,t)}{\partial x_j^2} = L_jC_j \frac{\partial^2 Q_m(x_j,t)}{\partial t^2},
\end{equation}
where $x_j \in [-l/8,l/8]$ is the position variable along section $j$, and $t$ is time. To find a solution for this equation for modes $a$ and $b$, we use the ansatz
\begin{equation}\label{eqn:Qab}
\begin{split}
    Q_a(x_j,t) = & \ A_j \sin \left( k_{aj} x_j + \phi_{aj} \right) \left[ a^\dagger e^{i\omega_a t} - a e^{-i\omega_a t} \right], \\
    Q_b(x_j,t) = & \ B_j \sin \left( k_{bj} x_j + \phi_{bj} \right) \left[ b^\dagger e^{i\omega_b t} - b e^{-i\omega_b t} \right],
\end{split}
\end{equation}
where $k_{mj}=\omega_m/v_j$ is the wave vector of mode $m$ along section $j$, $v_j=1/\sqrt{L_jC_j}$ is the phase velocity along section $j$, $A_j$ and $B_j$ are normalization amplitudes, and $\phi_{mj}$ is to be determined from the symmetry constraints. The current and voltage along the ring are related to the charge through
\begin{equation}
    I_{m}(x_j,t) = \frac{\partial Q_m(x_j,t)}{\partial t}, \quad
    V_m(x_j,t) = \frac{1}{C_j} \frac{\partial Q_m(x_j,t)}{\partial x_j}.
\end{equation}
By substituting for $Q_m$ from Eq.~(\ref{eqn:Qab}), we get the current and voltage for the two modes as
\begin{equation}
\begin{split}
    I_a(x_j,t) & = i\omega_a A_j \sin \left( k_{aj} x_j + \phi_{aj} \right) \left[ a^\dagger e^{i\omega_a t} + a e^{-i\omega_a t} \right], \\
    I_b(x_j,t) & = i\omega_b B_j \sin \left( k_{bj} x_j + \phi_{bj} \right) \left[ b^\dagger e^{i\omega_b t} + b e^{-i\omega_b t} \right], \\
    V_a(x_j,t) & = \omega_a Z_j A_j \cos \left( k_{aj} x_j + \phi_{aj} \right) \left[ a^\dagger e^{i\omega_a t} - a e^{-i\omega_a t} \right], \\
    V_b(x_j,t) & = \omega_b Z_j B_j \cos \left( k_{bj} x_j + \phi_{bj} \right) \left[ b^\dagger e^{i\omega_b t} - b e^{-i\omega_b t} \right],
\end{split}
\end{equation}
where $Z_j=\sqrt{L_j/C_j}$ is the characteristic impedance of section $j$.
Applying the symmetry constraints so that $V_a$ has antinodes at $x_{1/3}=0$ and nodes at $x_{2/4}=0$, we get the phase $\phi_{aj} = (j-1)\pi/2$. Similarly, for $V_b$ antinodes at $x_{2/4}=0$ and nodes at $x_{1/3}=0$, we get $\phi_{bj} = j\pi/2$.

Now, we apply the boundary conditions at the intersection point between sections 1 and 2 ({\it{i.e.,}} at $x_1=l/8$ and $x_2=-l/8$). For continuous current and voltage for mode $a$, we get
\begin{subequations}
\begin{equation}\label{eqn:BCIa}
    A_1 \sin\left( k_{a1} l/8 \right) = A_2 \sin\left( -k_{a2} l/8 + \pi/2\right),
\end{equation}
\begin{equation} \label{eqn:BCVa}
    A_1 Z_a \cos\left( k_{a1} l/8 \right) = A_2 Z_b \cos\left( -k_{a2} l/8 + \pi/2 \right).
\end{equation}
\end{subequations}
By dividing Eq.(\ref{eqn:BCVa}) by Eq.(\ref{eqn:BCIa}), substituting for $k_{a1}$ and $k_{a2}$, and rearranging, we reach the relation defining~$\omega_a$
\begin{equation}
    \tan\left( \frac{\omega_a}{v_a} \frac{l}{8} \right) \tan\left( \frac{\omega_a}{v_b} \frac{l}{8} \right) = \frac{Z_a}{Z_b}.
\end{equation}
Similarly, the conditions for continuous current and voltage for mode $b$ are
\begin{subequations}
\begin{equation}\label{eqn:BCIb}
    B_1 \sin\left( k_{b1} l/8 +\pi/2 \right) = B_2 \sin\left( -k_{b2} l/8 + \pi\right),
\end{equation}
\begin{equation}\label{eqn:BCVb}
    B_1 Z_a \cos\left( k_{b1} l/8 +\pi/2 \right) = B_2 Z_b \cos\left( -k_{b2} l/8 + \pi\right).
\end{equation}
\end{subequations}
Dividing Eq.(\ref{eqn:BCVb}) by Eq.(\ref{eqn:BCIb}), substituting for $k_{b1}$ and $k_{b2}$, and rearranging, we get $\omega_b$ from
\begin{equation}
    \tan\left( \frac{\omega_b}{v_a} \frac{l}{8} \right) \tan\left( \frac{\omega_b}{v_b} \frac{l}{8} \right) = \frac{Z_b}{Z_a}.
\end{equation}

\section{Frequency Shift by DC Current}\label{app:resIDC}
\begin{figure}
    \centering
    \includegraphics [width=0.8\linewidth] {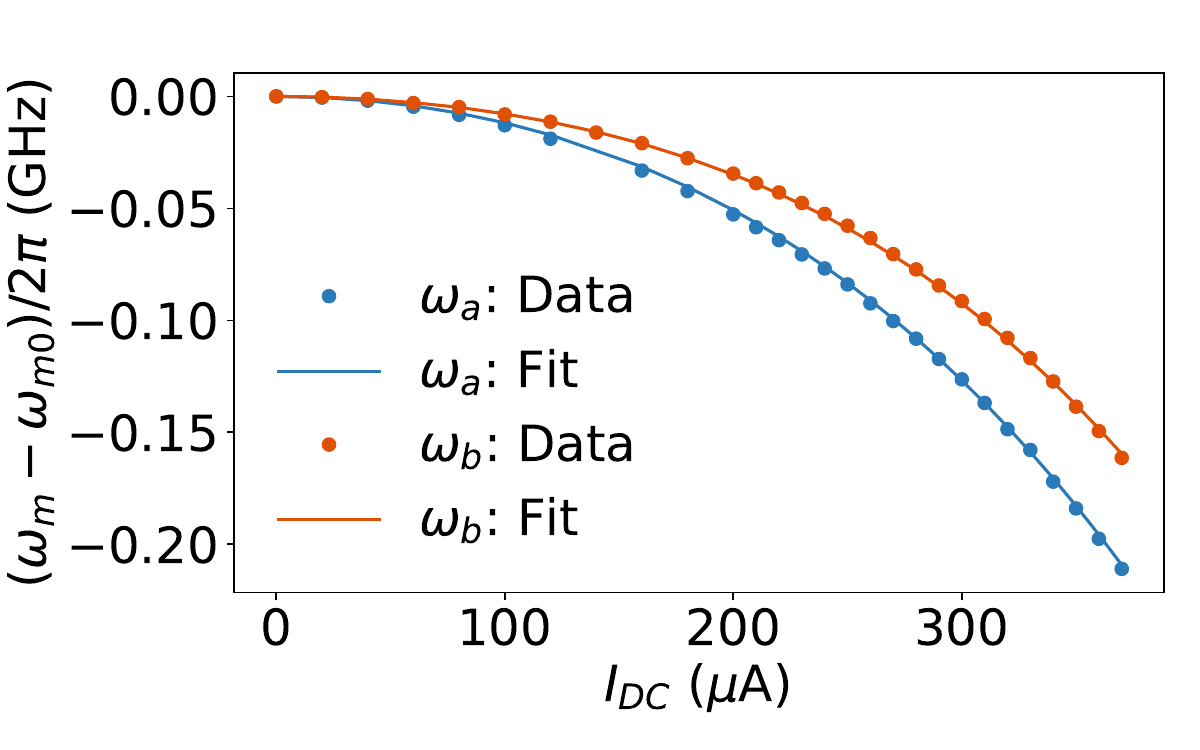}
    \caption{Dependence of the resonance frequency of the two modes on DC current. The fitting is done according to Eq.~(\ref{eqn:res}) with the current-dependent inductance in Eq.~(\ref{eqn:induct}).}
    \label{fig:tun}
\end{figure}
The DC current that provides the 3WM in the nanowire causes a shift in the device resonances as measured from the transmission spectrum at low input probe power, as shown in Fig.~\ref{fig:tun}. 
This occurs due to the self-consistent suppression of the condensate, and KI, with the supercurrent~\cite{clem2012kinetic}.
Assuming that the inductance is dominated by KI, the relation between the device resonances and the DC current can be obtained from Eq.~(\ref{eqn:res}) and the current-dependent expression of inductance
\begin{equation}\label{eqn:induct}
    L_a = L_b = L_0 \left[ 1 + \left( \frac{I_{\rm DC}/2}{I_*} \right)^2 + \alpha \left( \frac{I_{\rm DC}/2}{I_*} \right)^4 \right].
\end{equation}
Here $I_*$ is the effective nonlinear current scale which is proportional to the critical current of the nanowire, and $\alpha$ is a unitless coefficient of the higher-order nonlinearity. The value of $I_{\rm DC}$ is divided by 2 in Eq.~(\ref{eqn:induct}) because the current is divided into two branches in the device. By fitting the data in Fig.~\ref{fig:tun} to Eq.~(\ref{eqn:res}) with the current-dependent inductance in Eq.~(\ref{eqn:induct}), we extract $I_*=779$ $\mu$A and $1033$ $\mu$A for modes $a$ and $b$, respectively. This mismatch could be attributed to local defects in the nanowire or to a nonuniform width due to fabrication imperfections.

\section{Full Measurement Setup}
\label{app:setup}

\begin{figure}
    \centering
    \includegraphics [width=0.99\linewidth] {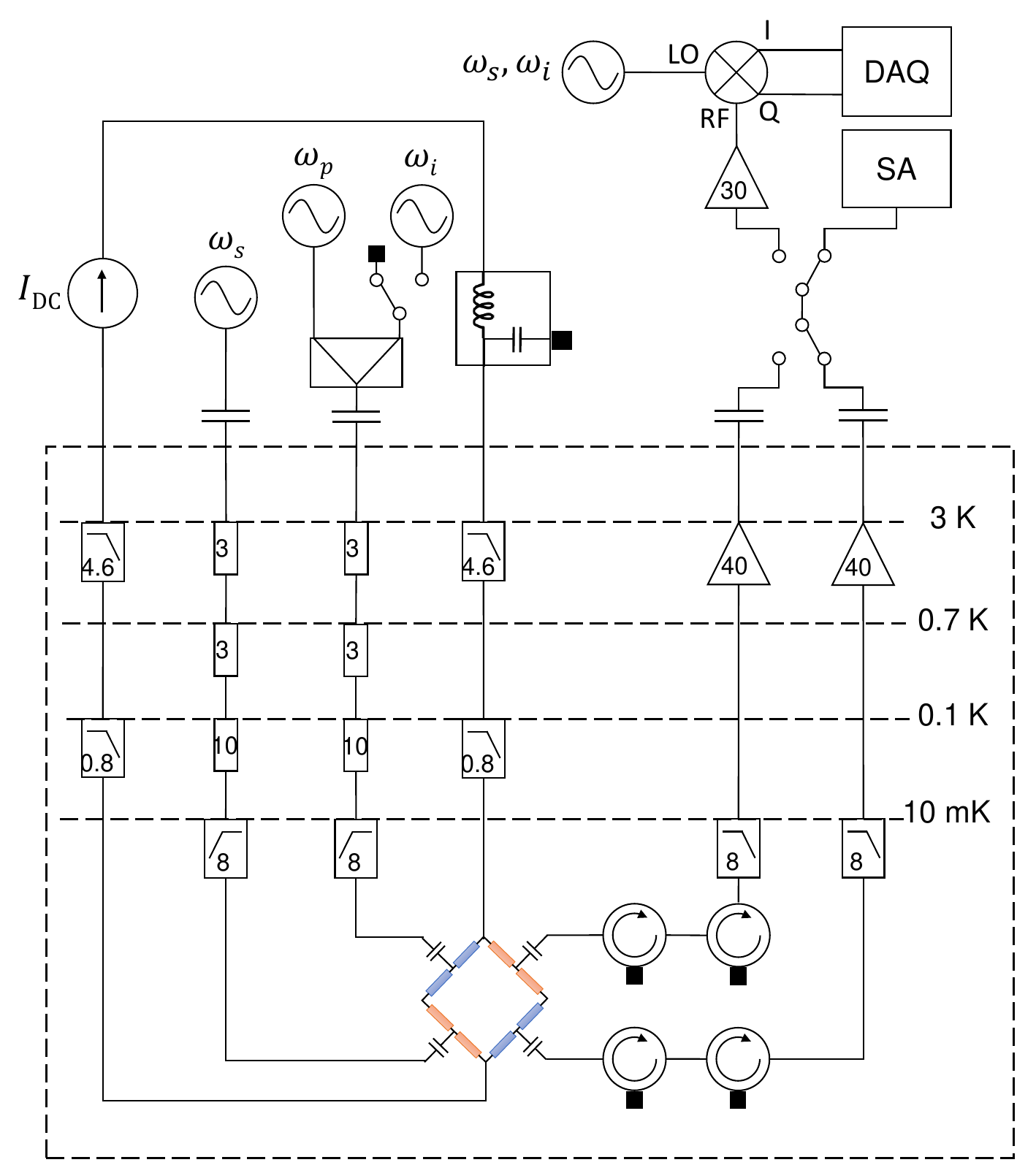}
    \caption{Full measurement setup. Attenuation and amplification values are shown in dB. The cutoff frequencies of the filters are shown in GHz. The black squares represent 50 $\Omega$ termination.}
    \label{fig:Setup}
\end{figure}

The full measurement setup is shown in Fig. \ref{fig:Setup}. The device is measured in a dilution refrigerator at 10 mK. The two input microwave lines are connected to ports 1 and 2 of the device. The input tones are generated by microwave sources at room temperature and attenuated inside the refrigerator. For the interference experiment, the idler and pump tones are combined and sent through the same input line. 
Each input line has a total of 16 dB attenuation distributed over different stages, as shown in Fig.~\ref{fig:Setup}, in addition to a high-pass filter (VHF-8400+) with 8 GHz cutoff frequency placed on the mixing chamber to filter the noise at the signal and idler frequencies while passing the pump. The high-pass filter has insertion loss of 36 dB at $\omega_a$ and 25 dB at $\omega_b$. 
The filters were installed on both input lines in order to enable the pump to be inserted either from the signal or idler side.
To estimate the attenuation of the filters at $\omega_a$ and $\omega_b$ at cryogenic temperatures, the filters were calibrated at 4~K separately.

We note that after adding these high-pass filters to the setup, several parasitic resonances arise in the transmission spectrum that do not exist without the filters. This is attributed to the formation of a short resonator between the filters and the device. This is not an issue, though, since the two modes of our device are easy to determine by observing their shift with DC current. An optimized version of the setup would avoid these filters by separating the pump into a third input line, then combining the pump and the idler (or signal) at the mixing chamber by a diplexer before going to the device.

The two output lines are connected to ports 3 and 4 of the device. On each line, the output goes through two circulators (LNF-CIC4\_8A) with a total of 40 dB isolation and is then filtered by a low-pass filter (VLF-7200+) of 8 GHz cutoff frequency to pass the signal and idler while reflecting the pump. High electron mobility transistor (HEMT) low-noise amplifiers (LNF-LNC4\_8C) are used to amplify the outputs at 3 K. Then, at room temperature, the outputs are either measured directly by a spectrum analyzer (SA), or are further amplified and down-converted to DC by a mixer (MMIQ-0218L) for measuring their quadratures via a data acquisition card (DAQ).

The DC current is generated by a current source through a series of low-pass filters; a bias-Tee (ZFBT-4R2G+) at room temperature, and two low-pass filters at 3 K (VLF-3800+) and 0.1 K (VLF-630+). The room-temperature low-pass filtering achieved by the bias-T with cutoff frequency $\sim10$ kHz was found to be important to realize stable 30 dB gain. Although slightly higher pump powers than shown in Fig.~\ref{fig:Amp}(d) and Fig.~\ref{fig:Amp}(e) produced gains approaching 50 dB, the data are not shown here since the device stability was an issue. This could be because of the device entering the bistability regime or because of the high sensitivity to DC current noise at high gain.

\section{Noise Performance} \label{app:noise}
\begin{figure}
    \centering
    \includegraphics [width=0.7\linewidth] {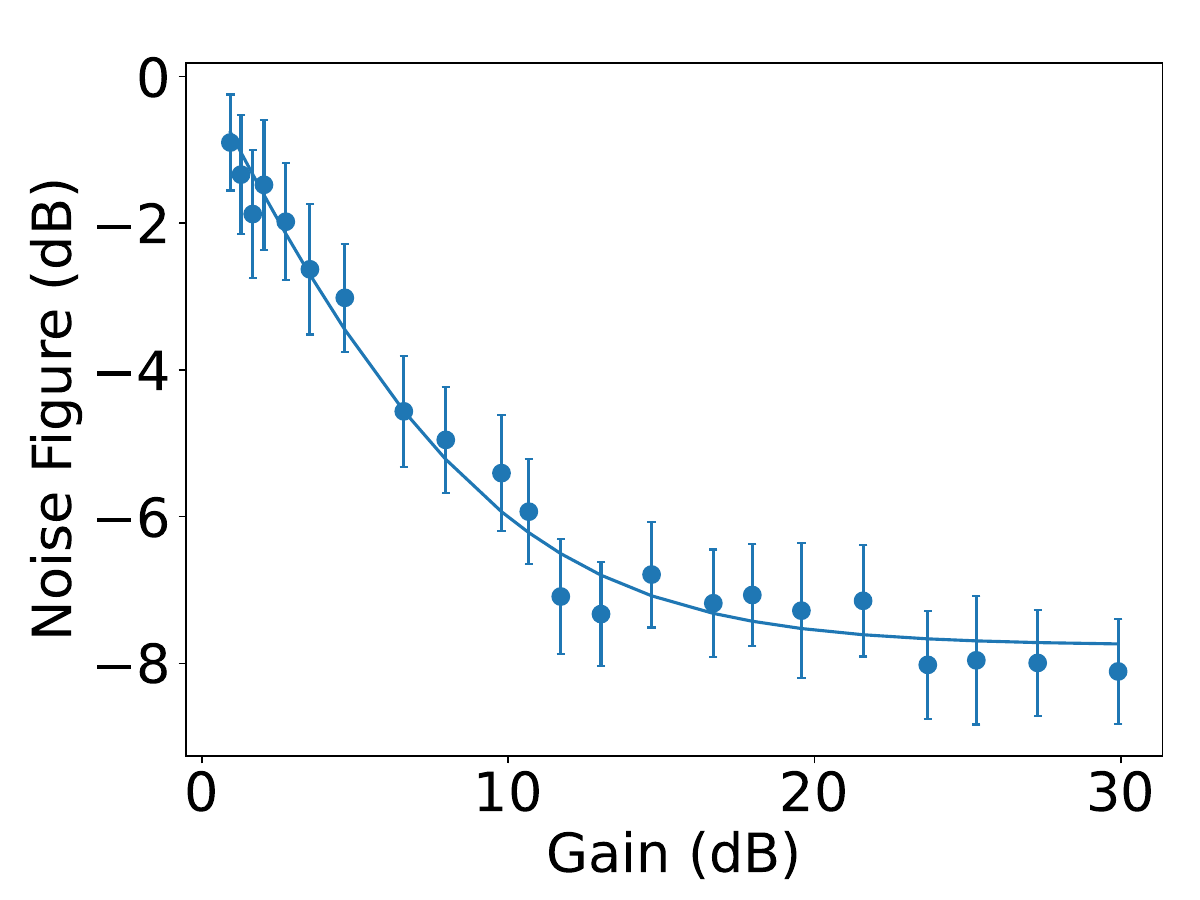}
    \caption{Noise figure versus signal gain. Solid line represents the data fitting to Eq.(\ref{eqn:NF}).}
    \label{fig:noise}
\end{figure}
The signal-to-noise ratio (SNR) improvement due to the KIPC is the reciprocal of the noise figure $\operatorname{NF}=\operatorname{SNR}_{\rm off} / \operatorname{SNR}_{\rm on}$, which we model by
\begin{equation}\label{eqn:NF}
\text{NF} = \frac{1}{G} \frac{N_{\rm sys} + (G-1) N_{\rm KIPC}}{N_{\rm sys}}.
\end{equation}
Here, $\text{SNR}_{\rm on}(G)$ is the SNR when the KIPC is pumped to get signal gain $G>1$, $\text{SNR}_{\rm off}$ is the SNR when the KIPC is not pumped so that $G=1$, $N_{\rm sys}$ is the noise power of the HEMT at 2.8 K in our measurement, and $N_{\rm KIPC}$ is the input referred noise power added by the KIPC. 
In Eq.(\ref{eqn:NF}), we assume zero input noise, which leads to an overestimation of $N_{\rm KIPC}$.
Figure \ref{fig:noise} shows the measured noise figure at the signal output versus signal gain. At small gain, the noise figure decreases linearly with gain. As the gain becomes comparable to the ratio $N_{\rm sys}/N_{\rm KIPC}$, the noise figure starts saturating. The measured data is fitted to Eq.~(\ref{eqn:NF}) to extract $N_{\rm KIPC}/N_{\rm sys} = 0.167(8)$, corresponding to SNR improvement of ${\sim8}$ dB. 
Assuming that $N_{\rm sys}$ is equivalent to 4.8 K noise temperature (2 K added noise by the HEMT as per the datasheet of the manufacturer plus 2.8 K thermal noise), which corresponds to $\sim$15.6 photons at 6.2 GHz, we obtain a rough estimate of $\sim$2.6 added noise photons by the KIPC. The estimate of 15.6 noise photons added by the HEMT agrees with the value obtained by the noise calibration of similar systems in literature \cite{mohamed2023selective}.

The ideal operation in the quantum-limited regime with only 0.5 added noise photons corresponds to a SNR improvement of 15 dB instead of the 8 dB obtained in Fig. \ref{fig:noise}. We expect that better noise performance could be achieved by optimizing our measurement setup. Improvements in thermal anchoring of the high-pass filters might help to reduce the noise injected into the circuit. Further reduction in added noise might be possible by separating the pump input line from the idler line, and adding filters to the pump line with higher rejection at the signal and idler frequencies.

\section{Input-Output Theory for the KIPC} \label{app:inout}
The Hamiltonian in Eq.~\eqref{eq:hamiltonian},
\begin{equation}
    H_r = \hbar \Delta_a a^\dagger a + \hbar \Delta_b b^\dagger b + \frac{\hbar}{2} \left( \xi a^\dagger b^\dagger + \xi^* a b \right),
\end{equation}
describes only the ring resonator itself. To include the two input and output lines, we introduce modes $c_j(\Delta)$ at port $j$ and frequency $\omega=\Delta+\omega_p/2$, and add the port Hamiltonian
\begin{equation}
    H_p = \hbar \sum_{j=1}^4\int\!\operatorname{d}\!\Delta\, \Delta c_j^\dagger(\Delta)c_j(\Delta).
\end{equation}
We model the capacitive coupling by
\begin{equation}
\begin{alignedat}{3}
    H_c &= i\hbar\sqrt{\frac{\kappa_a}{2\pi}}\int\!\operatorname{d}\!\Delta\,&&\left[\left(c_1^\dagger(\Delta)+c_3^\dagger(\Delta)\right) a\right.\\&&&\left.-a^\dagger \Big(c_1(\Delta)+c_3(\Delta)\Big)\right]\\
    &+  i\hbar\sqrt{\frac{\kappa_b}{2\pi}}\int\!\operatorname{d}\!\Delta\,&&\left[\left(c_2^\dagger(\Delta)+c_4^\dagger(\Delta)\right) b\right.\\&&&\left.-b^\dagger \Big(c_2(\Delta)+c_4(\Delta)\Big)\right].
\end{alignedat}
\end{equation}
Here, we assume that the coupling is perfectly mode selective but frequency independent. The overall Hamiltonian becomes $H=H_r+H_p+H_c$.

Input-output theory~\cite{GardinerCollett1985} introduces the input and output fields
\begin{equation}
    c_{j,\mathrm{in}/\mathrm{out}}(t)=\frac{1}{\sqrt{2\pi}}\int\!\operatorname{d}\!\Delta\,\operatorname{e}^{-i\Delta(t-t_{0/1})}c_{j,0/1}(\Delta),
\end{equation}
where $t_0<t<t_1$ and $c_{j,0/1}(\Delta)$ is the Heisenberg representation of $c_j(\Delta)$ at time $t_0$ or $t_1$, respectively. Applying the Fourier transformation $c(\Delta)=\frac{1}{\sqrt{2\pi}}\int\!\operatorname{d}\!t\,\operatorname{e}^{i\Delta t}c(t)$ shows that these fields directly correspond to the observables $c_{j,0/1}(\Delta)$,
\begin{equation}
    c_{j,\mathrm{in/out}}(\Delta)=\operatorname{e}^{i\Delta t_{0/1}}c_{j,0/1}(\Delta).
\end{equation}
In terms of input fields, the Heisenberg equations of motion for modes $a$ and $b$ become
\begin{align}\label{eq:heisenberg}
\begin{split}
    \partial_t a&=\frac{1}{i\hbar}[a,H_r]-\sqrt{\kappa_a}(c_{1,\mathrm{in}}+c_{3,\mathrm{in}})-\kappa_a a\\
    & = -i\Delta_a a-i\frac{\xi}{2}b^\dagger-\sqrt{\kappa_a}(c_{1,\mathrm{in}}+c_{3,\mathrm{in}})-\kappa_a a,\\
    \partial_t b &=\frac{1}{i\hbar}[b,H_r]-\sqrt{\kappa_b}(c_{2,\mathrm{in}}+c_{4,\mathrm{in}})-\kappa_b b\\
    &= -i\Delta_b b-i\frac{\xi}{2}a^\dagger-\sqrt{\kappa_b}(c_{2,\mathrm{in}}+c_{4,\mathrm{in}})-\kappa_b b.    
\end{split}
\end{align}
Input and output are related by
\begin{align}\label{eq:inout}
\begin{split}
    \sqrt{\kappa_a}a&=c_{1,\mathrm{out}}-c_{1,\mathrm{in}}=c_{3,\mathrm{out}}-c_{3,\mathrm{in}},\\
    \sqrt{\kappa_b}b&=c_{2,\mathrm{out}}-c_{2,\mathrm{in}}=c_{4,\mathrm{out}}-c_{4,\mathrm{in}}.
\end{split}
\end{align}
In Eqs.~\eqref{eq:heisenberg} and~\eqref{eq:inout}, all operators are evaluated at time $t$.

Our goal is to express $c_{3,\mathrm{out}}(\Delta)$ and $c_{4,\mathrm{out}}(\Delta)$ in terms of $c_{j,\mathrm{in}}(\Delta)$. We Fourier transform the Heisenberg Eqs.~\eqref{eq:heisenberg} and their adjoints, and use Eq.~\eqref{eq:inout} to express $a$ in terms of $c_{3,\mathrm{in/out}}$ and $b$ in terms of $c_{4,\mathrm{in/out}}$. This yields
\begin{align}\label{eq:outmatrix}
\begin{split}
	\begin{pmatrix}
	c_{3,\mathrm{out}} \\ c_{4,\mathrm{out}}^\dagger \\ c_{4,\mathrm{out}} \\ c_{3,\mathrm{out}}^\dagger
	\end{pmatrix}
	&=\begin{pmatrix}
	c_{3,\mathrm{in}} \\ c_{4,\mathrm{in}}^\dagger \\ c_{4,\mathrm{in}} \\ c_{3,\mathrm{in}}^\dagger
	\end{pmatrix}\\
        &-K
	\begin{pmatrix}
	A_{ab}^{-1} & 0\\
	0 & A_{ba}^{-1}
	\end{pmatrix}
	K
	\begin{pmatrix}
	c_{1,\mathrm{in}} +c_{3,\mathrm{in}} \\ c_{2,\mathrm{in}}^\dagger+c_{4,\mathrm{in}}^\dagger \\ c_{2,\mathrm{in}}+c_{4,\mathrm{in}} \\ c_{1,\mathrm{in}}^\dagger+c_{3,\mathrm{in}}^\dagger
	\end{pmatrix},
\end{split}
\end{align}
where $K=\operatorname{diag}(\sqrt{\kappa_a},\sqrt{\kappa_b},\sqrt{\kappa_b},\sqrt{\kappa_a})$, 
\begin{equation}
    A_{ab}=\begin{pmatrix}
	\kappa_a-i(\Delta-\Delta_a) & i\xi/2 \\
	-i\xi^*/2 & \kappa_b-i(\Delta+\Delta_b) \\
	\end{pmatrix},
\end{equation}
and $A_{ba}$ is $A_{ab}$ with exchanged indices $a$ and $b$. In Eq.~\eqref{eq:outmatrix}, $c_j$ abbreviates $c_j(\Delta)$, and $c^\dagger_j$ stands for $[c_j(-\Delta)]^\dagger$. Rows 1 and 3 of Eq.~\eqref{eq:outmatrix} evaluate to
\begin{align}
\begin{split}
    c_{3,\mathrm{out}}&=c_{3,\mathrm{in}}+g_{ii}(c_{1,\mathrm{in}}+c_{3,\mathrm{in}})+g_{si}(c_{2,\mathrm{in}} + c_{4,\mathrm{in}})^\dagger,\\
    c_{4,\mathrm{out}}&=c_{4,\mathrm{in}}+g_{ss}(c_{2,\mathrm{in}}+c_{4,\mathrm{in}})+g_{is}(c_{1,\mathrm{in}} +c_{3,\mathrm{in}})^\dagger.
\end{split}
\end{align}
For completeness, we add that
\begin{align}
\begin{split}
    c_{1,\mathrm{out}}&=c_{1,\mathrm{in}}+g_{ii}(c_{1,\mathrm{in}}+c_{3,\mathrm{in}})+g_{si}(c_{2,\mathrm{in}} + c_{4,\mathrm{in}})^\dagger,\\
    c_{2,\mathrm{out}}&=c_{2,\mathrm{in}}+g_{ss}(c_{2,\mathrm{in}}+c_{4,\mathrm{in}})+g_{is}(c_{1,\mathrm{in}} +c_{3,\mathrm{in}})^\dagger.
\end{split}
\end{align}
The amplitude gains are
\begin{align}
\begin{split}
    g_{ss}&=\frac{\kappa_b}{D_s}[i(\Delta+\Delta_a)-\kappa_a]\\
    g_{si}&=i\frac{\sqrt{\kappa_a\kappa_b}}{D_i}\frac{\xi}{2}\\
    g_{ii}&=\frac{\kappa_a}{D_i}[i(\Delta+\Delta_b)-\kappa_b]\\
    g_{is}&=i\frac{\sqrt{\kappa_a\kappa_b}}{D_s}\frac{\xi}{2}
\end{split}
\end{align}
with 
\begin{equation}
    D_{s/i}=[i(\Delta\pm\Delta_a)-\kappa_a][i(\Delta\mp\Delta_b)-\kappa_b]-\frac{|\xi|^2}{4}.
\end{equation}
Note that we identify mode $b$, input port 2, and output port 4 as the signal, and mode $a$, input port 1, and output port 3 as the idler. The first (second) index of amplitude gain $g$ indicates the input (output).

Let us now specify to the three scenarios in Section~\ref{sec:results} of the main text. When measuring at signal frequency $\omega_s=\delta+\omega_p/2$ and idler frequency $\omega_i=-\delta+\omega_p/2$, the relevant amplitude gains become $g_{ss}(\delta)$, $g_{is}(\delta)$, $g_{ii}(-\delta)$, and $g_{si}(-\delta)$. Assuming that, at time $t_0$, the input fields are in the coherent eigenstates $|\alpha\rangle$ and $|\beta\rangle$ of $c_1(-\delta)$ and $c_2(\delta)$, respectively, and the output fields are in the vacuum state, input-output theory predicts the following measurement outcomes for the quadratures at the output ports:
\begin{align}
\begin{split}
    I_i+iQ_i&=\operatorname{e}^{i\delta(t_1-t_0)}[g_{ii}(-\delta)\alpha+g_{si}(-\delta)\beta^*]\\
    I_s+iQ_s&=\operatorname{e}^{-i\delta(t_1-t_0)}[g_{ss}(\delta)\beta+g_{is}(\delta)\alpha^*]
\end{split}
\end{align}
In the absence of idler input, $\alpha=0$, the relevant gains are further reduced to $g_{s}(\delta)=g_{ss}(\delta)$ and $g_i(-\delta)=g_{si}(-\delta)$, cf. Eq.~\eqref{eqn:gain} of the main text, and the power gains become
\begin{equation}
    G_{s/i}=\frac{|I_{s/i}|^2+|Q_{s/i}|^2}{|\beta|^2}=|g_{s/i}(\pm\delta)|^2.
\end{equation}
We observe that $I_s+iQ_s\propto \beta$ and $I_i+iQ_i\propto \beta^*$, which implies the phase correlations shown in Fig.~\ref{fig:Phase} of the main text (note that the phases of $\beta$ and the input signal have opposite sign). In the case of signal and idler input, complete extinction by destructive interference requires
\begin{equation}
\begin{alignedat}{3}
    \frac{\beta}{\alpha^*}&=-i\sqrt{\frac{\kappa_a}{\kappa_b}}\frac{\xi}{2[i(\delta+\Delta_a)-\kappa_a]}\quad&&\text{for zero signal output},\\
    \frac{\beta}{\alpha^*}&=-i\sqrt{\frac{\kappa_a}{\kappa_b}}\frac{2[i(\delta-\Delta_b)-\kappa_b]}{\xi^*}\quad&&\text{for zero idler output.}
\end{alignedat}
\end{equation}


\begin{thebibliography}{38}%
\makeatletter
\providecommand \@ifxundefined [1]{%
 \@ifx{#1\undefined}
}%
\providecommand \@ifnum [1]{%
 \ifnum #1\expandafter \@firstoftwo
 \else \expandafter \@secondoftwo
 \fi
}%
\providecommand \@ifx [1]{%
 \ifx #1\expandafter \@firstoftwo
 \else \expandafter \@secondoftwo
 \fi
}%
\providecommand \natexlab [1]{#1}%
\providecommand \enquote  [1]{``#1''}%
\providecommand \bibnamefont  [1]{#1}%
\providecommand \bibfnamefont [1]{#1}%
\providecommand \citenamefont [1]{#1}%
\providecommand \href@noop [0]{\@secondoftwo}%
\providecommand \href [0]{\begingroup \@sanitize@url \@href}%
\providecommand \@href[1]{\@@startlink{#1}\@@href}%
\providecommand \@@href[1]{\endgroup#1\@@endlink}%
\providecommand \@sanitize@url [0]{\catcode `\\12\catcode `\$12\catcode
  `\&12\catcode `\#12\catcode `\^12\catcode `\_12\catcode `\%12\relax}%
\providecommand \@@startlink[1]{}%
\providecommand \@@endlink[0]{}%
\providecommand \url  [0]{\begingroup\@sanitize@url \@url }%
\providecommand \@url [1]{\endgroup\@href {#1}{\urlprefix }}%
\providecommand \urlprefix  [0]{URL }%
\providecommand \Eprint [0]{\href }%
\providecommand \doibase [0]{https://doi.org/}%
\providecommand \selectlanguage [0]{\@gobble}%
\providecommand \bibinfo  [0]{\@secondoftwo}%
\providecommand \bibfield  [0]{\@secondoftwo}%
\providecommand \translation [1]{[#1]}%
\providecommand \BibitemOpen [0]{}%
\providecommand \bibitemStop [0]{}%
\providecommand \bibitemNoStop [0]{.\EOS\space}%
\providecommand \EOS [0]{\spacefactor3000\relax}%
\providecommand \BibitemShut  [1]{\csname bibitem#1\endcsname}%
\let\auto@bib@innerbib\@empty
\bibitem [{\citenamefont {Ofek}\ \emph {et~al.}(2016)\citenamefont {Ofek},
  \citenamefont {Petrenko}, \citenamefont {Heeres}, \citenamefont {Reinhold},
  \citenamefont {Leghtas}, \citenamefont {Vlastakis}, \citenamefont {Liu},
  \citenamefont {Frunzio}, \citenamefont {Girvin}, \citenamefont {Jiang} \emph
  {et~al.}}]{ofek2016extending}%
  \BibitemOpen
  \bibfield  {author} {\bibinfo {author} {\bibfnamefont {N.}~\bibnamefont
  {Ofek}}, \bibinfo {author} {\bibfnamefont {A.}~\bibnamefont {Petrenko}},
  \bibinfo {author} {\bibfnamefont {R.}~\bibnamefont {Heeres}}, \bibinfo
  {author} {\bibfnamefont {P.}~\bibnamefont {Reinhold}}, \bibinfo {author}
  {\bibfnamefont {Z.}~\bibnamefont {Leghtas}}, \bibinfo {author} {\bibfnamefont
  {B.}~\bibnamefont {Vlastakis}}, \bibinfo {author} {\bibfnamefont
  {Y.}~\bibnamefont {Liu}}, \bibinfo {author} {\bibfnamefont {L.}~\bibnamefont
  {Frunzio}}, \bibinfo {author} {\bibfnamefont {S.}~\bibnamefont {Girvin}},
  \bibinfo {author} {\bibfnamefont {L.}~\bibnamefont {Jiang}}, \emph {et~al.},\
  }\bibfield  {title} {\bibinfo {title} {Extending the lifetime of a quantum
  bit with error correction in superconducting circuits},\ }\href
  {https://doi.org/10.1038/nature18949} {\bibfield  {journal} {\bibinfo
  {journal} {Nature}\ }\textbf {\bibinfo {volume} {536}},\ \bibinfo {pages}
  {441} (\bibinfo {year} {2016})}\BibitemShut {NoStop}%
\bibitem [{\citenamefont {Chapman}\ \emph {et~al.}(2023)\citenamefont
  {Chapman}, \citenamefont {de~Graaf}, \citenamefont {Xue}, \citenamefont
  {Zhang}, \citenamefont {Teoh}, \citenamefont {Curtis}, \citenamefont
  {Tsunoda}, \citenamefont {Eickbusch}, \citenamefont {Read}, \citenamefont
  {Koottandavida} \emph {et~al.}}]{chapman2023high}%
  \BibitemOpen
  \bibfield  {author} {\bibinfo {author} {\bibfnamefont {B.~J.}\ \bibnamefont
  {Chapman}}, \bibinfo {author} {\bibfnamefont {S.~J.}\ \bibnamefont
  {de~Graaf}}, \bibinfo {author} {\bibfnamefont {S.~H.}\ \bibnamefont {Xue}},
  \bibinfo {author} {\bibfnamefont {Y.}~\bibnamefont {Zhang}}, \bibinfo
  {author} {\bibfnamefont {J.}~\bibnamefont {Teoh}}, \bibinfo {author}
  {\bibfnamefont {J.~C.}\ \bibnamefont {Curtis}}, \bibinfo {author}
  {\bibfnamefont {T.}~\bibnamefont {Tsunoda}}, \bibinfo {author} {\bibfnamefont
  {A.}~\bibnamefont {Eickbusch}}, \bibinfo {author} {\bibfnamefont {A.~P.}\
  \bibnamefont {Read}}, \bibinfo {author} {\bibfnamefont {A.}~\bibnamefont
  {Koottandavida}}, \emph {et~al.},\ }\bibfield  {title} {\bibinfo {title}
  {High-on-off-ratio beam-splitter interaction for gates on bosonically encoded
  qubits},\ }\href {https://doi.org/10.1103/PRXQuantum.4.020355} {\bibfield
  {journal} {\bibinfo  {journal} {PRX Quantum}\ }\textbf {\bibinfo {volume}
  {4}},\ \bibinfo {pages} {020355} (\bibinfo {year} {2023})}\BibitemShut
  {NoStop}%
\bibitem [{\citenamefont {He}\ \emph {et~al.}(2023)\citenamefont {He},
  \citenamefont {Lu}, \citenamefont {Bao}, \citenamefont {Xue}, \citenamefont
  {Jiang}, \citenamefont {Wang}, \citenamefont {Roudsari}, \citenamefont
  {Delsing}, \citenamefont {Tsai},\ and\ \citenamefont {Lin}}]{he2023fast}%
  \BibitemOpen
  \bibfield  {author} {\bibinfo {author} {\bibfnamefont {X.}~\bibnamefont
  {He}}, \bibinfo {author} {\bibfnamefont {Y.}~\bibnamefont {Lu}}, \bibinfo
  {author} {\bibfnamefont {D.}~\bibnamefont {Bao}}, \bibinfo {author}
  {\bibfnamefont {H.}~\bibnamefont {Xue}}, \bibinfo {author} {\bibfnamefont
  {W.}~\bibnamefont {Jiang}}, \bibinfo {author} {\bibfnamefont
  {Z.}~\bibnamefont {Wang}}, \bibinfo {author} {\bibfnamefont {A.}~\bibnamefont
  {Roudsari}}, \bibinfo {author} {\bibfnamefont {P.}~\bibnamefont {Delsing}},
  \bibinfo {author} {\bibfnamefont {J.}~\bibnamefont {Tsai}},\ and\ \bibinfo
  {author} {\bibfnamefont {Z.}~\bibnamefont {Lin}},\ }\bibfield  {title}
  {\bibinfo {title} {Fast generation of schr{\"o}dinger cat states using a
  kerr-tunable superconducting resonator},\ }\href
  {https://doi.org/https://doi.org/10.1038/s41467-023-42057-0} {\bibfield
  {journal} {\bibinfo  {journal} {Nature communications}\ }\textbf {\bibinfo
  {volume} {14}},\ \bibinfo {pages} {6358} (\bibinfo {year}
  {2023})}\BibitemShut {NoStop}%
\bibitem [{\citenamefont {Pogorzalek}\ \emph {et~al.}(2019)\citenamefont
  {Pogorzalek}, \citenamefont {Fedorov}, \citenamefont {Xu}, \citenamefont
  {Parra-Rodriguez}, \citenamefont {Sanz}, \citenamefont {Fischer},
  \citenamefont {Xie}, \citenamefont {Inomata}, \citenamefont {Nakamura},
  \citenamefont {Solano} \emph {et~al.}}]{pogorzalek2019secure}%
  \BibitemOpen
  \bibfield  {author} {\bibinfo {author} {\bibfnamefont {S.}~\bibnamefont
  {Pogorzalek}}, \bibinfo {author} {\bibfnamefont {K.}~\bibnamefont {Fedorov}},
  \bibinfo {author} {\bibfnamefont {M.}~\bibnamefont {Xu}}, \bibinfo {author}
  {\bibfnamefont {A.}~\bibnamefont {Parra-Rodriguez}}, \bibinfo {author}
  {\bibfnamefont {M.}~\bibnamefont {Sanz}}, \bibinfo {author} {\bibfnamefont
  {M.}~\bibnamefont {Fischer}}, \bibinfo {author} {\bibfnamefont
  {E.}~\bibnamefont {Xie}}, \bibinfo {author} {\bibfnamefont {K.}~\bibnamefont
  {Inomata}}, \bibinfo {author} {\bibfnamefont {Y.}~\bibnamefont {Nakamura}},
  \bibinfo {author} {\bibfnamefont {E.}~\bibnamefont {Solano}}, \emph
  {et~al.},\ }\bibfield  {title} {\bibinfo {title} {Secure quantum remote state
  preparation of squeezed microwave states},\ }\href
  {https://doi.org/10.1038/s41467-019-10727-7} {\bibfield  {journal} {\bibinfo
  {journal} {Nature communications}\ }\textbf {\bibinfo {volume} {10}},\
  \bibinfo {pages} {1} (\bibinfo {year} {2019})}\BibitemShut {NoStop}%
\bibitem [{\citenamefont {Fedorov}\ \emph {et~al.}(2021)\citenamefont
  {Fedorov}, \citenamefont {Renger}, \citenamefont {Pogorzalek}, \citenamefont
  {Di~Candia}, \citenamefont {Chen}, \citenamefont {Nojiri}, \citenamefont
  {Inomata}, \citenamefont {Nakamura}, \citenamefont {Partanen}, \citenamefont
  {Marx} \emph {et~al.}}]{fedorov2021experimental}%
  \BibitemOpen
  \bibfield  {author} {\bibinfo {author} {\bibfnamefont {K.~G.}\ \bibnamefont
  {Fedorov}}, \bibinfo {author} {\bibfnamefont {M.}~\bibnamefont {Renger}},
  \bibinfo {author} {\bibfnamefont {S.}~\bibnamefont {Pogorzalek}}, \bibinfo
  {author} {\bibfnamefont {R.}~\bibnamefont {Di~Candia}}, \bibinfo {author}
  {\bibfnamefont {Q.}~\bibnamefont {Chen}}, \bibinfo {author} {\bibfnamefont
  {Y.}~\bibnamefont {Nojiri}}, \bibinfo {author} {\bibfnamefont
  {K.}~\bibnamefont {Inomata}}, \bibinfo {author} {\bibfnamefont
  {Y.}~\bibnamefont {Nakamura}}, \bibinfo {author} {\bibfnamefont
  {M.}~\bibnamefont {Partanen}}, \bibinfo {author} {\bibfnamefont
  {A.}~\bibnamefont {Marx}}, \emph {et~al.},\ }\bibfield  {title} {\bibinfo
  {title} {Experimental quantum teleportation of propagating microwaves},\
  }\href {https://doi.org/https://doi.org/10.1126/sciadv.abk0891} {\bibfield
  {journal} {\bibinfo  {journal} {Science advances}\ }\textbf {\bibinfo
  {volume} {7}},\ \bibinfo {pages} {eabk0891} (\bibinfo {year}
  {2021})}\BibitemShut {NoStop}%
\bibitem [{\citenamefont {Silveri}\ \emph {et~al.}(2016)\citenamefont
  {Silveri}, \citenamefont {Zalys-Geller}, \citenamefont {Hatridge},
  \citenamefont {Leghtas}, \citenamefont {Devoret},\ and\ \citenamefont
  {Girvin}}]{silveri2016theory}%
  \BibitemOpen
  \bibfield  {author} {\bibinfo {author} {\bibfnamefont {M.}~\bibnamefont
  {Silveri}}, \bibinfo {author} {\bibfnamefont {E.}~\bibnamefont
  {Zalys-Geller}}, \bibinfo {author} {\bibfnamefont {M.}~\bibnamefont
  {Hatridge}}, \bibinfo {author} {\bibfnamefont {Z.}~\bibnamefont {Leghtas}},
  \bibinfo {author} {\bibfnamefont {M.~H.}\ \bibnamefont {Devoret}},\ and\
  \bibinfo {author} {\bibfnamefont {S.~M.}\ \bibnamefont {Girvin}},\ }\bibfield
   {title} {\bibinfo {title} {Theory of remote entanglement via quantum-limited
  phase-preserving amplification},\ }\href
  {https://doi.org/10.1103/PhysRevA.93.062310} {\bibfield  {journal} {\bibinfo
  {journal} {Physical Review A}\ }\textbf {\bibinfo {volume} {93}},\ \bibinfo
  {pages} {062310} (\bibinfo {year} {2016})}\BibitemShut {NoStop}%
\bibitem [{\citenamefont {Barzanjeh}\ \emph {et~al.}(2020)\citenamefont
  {Barzanjeh}, \citenamefont {Pirandola}, \citenamefont {Vitali},\ and\
  \citenamefont {Fink}}]{barzanjeh2020microwave}%
  \BibitemOpen
  \bibfield  {author} {\bibinfo {author} {\bibfnamefont {S.}~\bibnamefont
  {Barzanjeh}}, \bibinfo {author} {\bibfnamefont {S.}~\bibnamefont
  {Pirandola}}, \bibinfo {author} {\bibfnamefont {D.}~\bibnamefont {Vitali}},\
  and\ \bibinfo {author} {\bibfnamefont {J.~M.}\ \bibnamefont {Fink}},\
  }\bibfield  {title} {\bibinfo {title} {Microwave quantum illumination using a
  digital receiver},\ }\href
  {https://doi.org/https://doi.org/10.1126/sciadv.abb0451} {\bibfield
  {journal} {\bibinfo  {journal} {Science advances}\ }\textbf {\bibinfo
  {volume} {6}},\ \bibinfo {pages} {eabb0451} (\bibinfo {year}
  {2020})}\BibitemShut {NoStop}%
\bibitem [{\citenamefont {Flurin}\ \emph {et~al.}(2012)\citenamefont {Flurin},
  \citenamefont {Roch}, \citenamefont {Mallet}, \citenamefont {Devoret},\ and\
  \citenamefont {Huard}}]{flurin2012generating}%
  \BibitemOpen
  \bibfield  {author} {\bibinfo {author} {\bibfnamefont {E.}~\bibnamefont
  {Flurin}}, \bibinfo {author} {\bibfnamefont {N.}~\bibnamefont {Roch}},
  \bibinfo {author} {\bibfnamefont {F.}~\bibnamefont {Mallet}}, \bibinfo
  {author} {\bibfnamefont {M.~H.}\ \bibnamefont {Devoret}},\ and\ \bibinfo
  {author} {\bibfnamefont {B.}~\bibnamefont {Huard}},\ }\bibfield  {title}
  {\bibinfo {title} {Generating entangled microwave radiation over two
  transmission lines},\ }\href
  {https://doi.org/https://doi.org/10.1103/PhysRevLett.109.183901} {\bibfield
  {journal} {\bibinfo  {journal} {Physical review letters}\ }\textbf {\bibinfo
  {volume} {109}},\ \bibinfo {pages} {183901} (\bibinfo {year}
  {2012})}\BibitemShut {NoStop}%
\bibitem [{\citenamefont {Bergeal}\ \emph
  {et~al.}(2010{\natexlab{a}})\citenamefont {Bergeal}, \citenamefont {Vijay},
  \citenamefont {Manucharyan}, \citenamefont {Siddiqi}, \citenamefont
  {Schoelkopf}, \citenamefont {Girvin},\ and\ \citenamefont
  {Devoret}}]{bergeal2010analog}%
  \BibitemOpen
  \bibfield  {author} {\bibinfo {author} {\bibfnamefont {N.}~\bibnamefont
  {Bergeal}}, \bibinfo {author} {\bibfnamefont {R.}~\bibnamefont {Vijay}},
  \bibinfo {author} {\bibfnamefont {V.}~\bibnamefont {Manucharyan}}, \bibinfo
  {author} {\bibfnamefont {I.}~\bibnamefont {Siddiqi}}, \bibinfo {author}
  {\bibfnamefont {R.}~\bibnamefont {Schoelkopf}}, \bibinfo {author}
  {\bibfnamefont {S.}~\bibnamefont {Girvin}},\ and\ \bibinfo {author}
  {\bibfnamefont {M.}~\bibnamefont {Devoret}},\ }\bibfield  {title} {\bibinfo
  {title} {Analog information processing at the quantum limit with a josephson
  ring modulator},\ }\href {https://doi.org/https://doi.org/10.1038/nphys1516}
  {\bibfield  {journal} {\bibinfo  {journal} {Nature Physics}\ }\textbf
  {\bibinfo {volume} {6}},\ \bibinfo {pages} {296} (\bibinfo {year}
  {2010}{\natexlab{a}})}\BibitemShut {NoStop}%
\bibitem [{\citenamefont {Abdo}\ \emph
  {et~al.}(2013{\natexlab{a}})\citenamefont {Abdo}, \citenamefont {Kamal},\
  and\ \citenamefont {Devoret}}]{abdo2013nondegenerate}%
  \BibitemOpen
  \bibfield  {author} {\bibinfo {author} {\bibfnamefont {B.}~\bibnamefont
  {Abdo}}, \bibinfo {author} {\bibfnamefont {A.}~\bibnamefont {Kamal}},\ and\
  \bibinfo {author} {\bibfnamefont {M.}~\bibnamefont {Devoret}},\ }\bibfield
  {title} {\bibinfo {title} {Nondegenerate three-wave mixing with the josephson
  ring modulator},\ }\href
  {https://doi.org/https://doi.org/10.1103/PhysRevB.87.014508} {\bibfield
  {journal} {\bibinfo  {journal} {Physical Review B}\ }\textbf {\bibinfo
  {volume} {87}},\ \bibinfo {pages} {014508} (\bibinfo {year}
  {2013}{\natexlab{a}})}\BibitemShut {NoStop}%
\bibitem [{\citenamefont {Abdo}\ \emph
  {et~al.}(2013{\natexlab{b}})\citenamefont {Abdo}, \citenamefont {Sliwa},
  \citenamefont {Schackert}, \citenamefont {Bergeal}, \citenamefont {Hatridge},
  \citenamefont {Frunzio}, \citenamefont {Stone},\ and\ \citenamefont
  {Devoret}}]{abdo2013full}%
  \BibitemOpen
  \bibfield  {author} {\bibinfo {author} {\bibfnamefont {B.}~\bibnamefont
  {Abdo}}, \bibinfo {author} {\bibfnamefont {K.}~\bibnamefont {Sliwa}},
  \bibinfo {author} {\bibfnamefont {F.}~\bibnamefont {Schackert}}, \bibinfo
  {author} {\bibfnamefont {N.}~\bibnamefont {Bergeal}}, \bibinfo {author}
  {\bibfnamefont {M.}~\bibnamefont {Hatridge}}, \bibinfo {author}
  {\bibfnamefont {L.}~\bibnamefont {Frunzio}}, \bibinfo {author} {\bibfnamefont
  {A.~D.}\ \bibnamefont {Stone}},\ and\ \bibinfo {author} {\bibfnamefont
  {M.}~\bibnamefont {Devoret}},\ }\bibfield  {title} {\bibinfo {title} {Full
  coherent frequency conversion between two propagating microwave modes},\
  }\href {https://doi.org/https://doi.org/10.1103/PhysRevLett.110.173902}
  {\bibfield  {journal} {\bibinfo  {journal} {Physical review letters}\
  }\textbf {\bibinfo {volume} {110}},\ \bibinfo {pages} {173902} (\bibinfo
  {year} {2013}{\natexlab{b}})}\BibitemShut {NoStop}%
\bibitem [{\citenamefont {Sliwa}\ \emph {et~al.}(2015)\citenamefont {Sliwa},
  \citenamefont {Hatridge}, \citenamefont {Narla}, \citenamefont {Shankar},
  \citenamefont {Frunzio}, \citenamefont {Schoelkopf},\ and\ \citenamefont
  {Devoret}}]{sliwa2015reconfigurable}%
  \BibitemOpen
  \bibfield  {author} {\bibinfo {author} {\bibfnamefont {K.}~\bibnamefont
  {Sliwa}}, \bibinfo {author} {\bibfnamefont {M.}~\bibnamefont {Hatridge}},
  \bibinfo {author} {\bibfnamefont {A.}~\bibnamefont {Narla}}, \bibinfo
  {author} {\bibfnamefont {S.}~\bibnamefont {Shankar}}, \bibinfo {author}
  {\bibfnamefont {L.}~\bibnamefont {Frunzio}}, \bibinfo {author} {\bibfnamefont
  {R.}~\bibnamefont {Schoelkopf}},\ and\ \bibinfo {author} {\bibfnamefont
  {M.}~\bibnamefont {Devoret}},\ }\bibfield  {title} {\bibinfo {title}
  {Reconfigurable josephson circulator/directional amplifier},\ }\href
  {https://doi.org/https://doi.org/10.1103/PhysRevX.5.041020} {\bibfield
  {journal} {\bibinfo  {journal} {Physical Review X}\ }\textbf {\bibinfo
  {volume} {5}},\ \bibinfo {pages} {041020} (\bibinfo {year}
  {2015})}\BibitemShut {NoStop}%
\bibitem [{\citenamefont {Abdo}\ \emph {et~al.}(2017)\citenamefont {Abdo},
  \citenamefont {Brink},\ and\ \citenamefont {Chow}}]{abdo2017gyrator}%
  \BibitemOpen
  \bibfield  {author} {\bibinfo {author} {\bibfnamefont {B.}~\bibnamefont
  {Abdo}}, \bibinfo {author} {\bibfnamefont {M.}~\bibnamefont {Brink}},\ and\
  \bibinfo {author} {\bibfnamefont {J.~M.}\ \bibnamefont {Chow}},\ }\bibfield
  {title} {\bibinfo {title} {Gyrator operation using josephson mixers},\ }\href
  {https://doi.org/https://doi.org/10.1103/PhysRevApplied.8.034009} {\bibfield
  {journal} {\bibinfo  {journal} {Physical Review Applied}\ }\textbf {\bibinfo
  {volume} {8}},\ \bibinfo {pages} {034009} (\bibinfo {year}
  {2017})}\BibitemShut {NoStop}%
\bibitem [{\citenamefont {Chien}\ \emph {et~al.}(2020)\citenamefont {Chien},
  \citenamefont {Lanes}, \citenamefont {Liu}, \citenamefont {Cao},
  \citenamefont {Lu}, \citenamefont {Motz}, \citenamefont {Liu}, \citenamefont
  {Pekker},\ and\ \citenamefont {Hatridge}}]{chien2020multiparametric}%
  \BibitemOpen
  \bibfield  {author} {\bibinfo {author} {\bibfnamefont {T.-C.}\ \bibnamefont
  {Chien}}, \bibinfo {author} {\bibfnamefont {O.}~\bibnamefont {Lanes}},
  \bibinfo {author} {\bibfnamefont {C.}~\bibnamefont {Liu}}, \bibinfo {author}
  {\bibfnamefont {X.}~\bibnamefont {Cao}}, \bibinfo {author} {\bibfnamefont
  {P.}~\bibnamefont {Lu}}, \bibinfo {author} {\bibfnamefont {S.}~\bibnamefont
  {Motz}}, \bibinfo {author} {\bibfnamefont {G.}~\bibnamefont {Liu}}, \bibinfo
  {author} {\bibfnamefont {D.}~\bibnamefont {Pekker}},\ and\ \bibinfo {author}
  {\bibfnamefont {M.}~\bibnamefont {Hatridge}},\ }\bibfield  {title} {\bibinfo
  {title} {Multiparametric amplification and qubit measurement with a kerr-free
  josephson ring modulator},\ }\href
  {https://doi.org/https://doi.org/10.1103/PhysRevA.101.042336} {\bibfield
  {journal} {\bibinfo  {journal} {Physical Review A}\ }\textbf {\bibinfo
  {volume} {101}},\ \bibinfo {pages} {042336} (\bibinfo {year}
  {2020})}\BibitemShut {NoStop}%
\bibitem [{\citenamefont {Hatridge}\ \emph {et~al.}(2013)\citenamefont
  {Hatridge}, \citenamefont {Shankar}, \citenamefont {Mirrahimi}, \citenamefont
  {Schackert}, \citenamefont {Geerlings}, \citenamefont {Brecht}, \citenamefont
  {Sliwa}, \citenamefont {Abdo}, \citenamefont {Frunzio}, \citenamefont
  {Girvin} \emph {et~al.}}]{hatridge2013quantum}%
  \BibitemOpen
  \bibfield  {author} {\bibinfo {author} {\bibfnamefont {M.}~\bibnamefont
  {Hatridge}}, \bibinfo {author} {\bibfnamefont {S.}~\bibnamefont {Shankar}},
  \bibinfo {author} {\bibfnamefont {M.}~\bibnamefont {Mirrahimi}}, \bibinfo
  {author} {\bibfnamefont {F.}~\bibnamefont {Schackert}}, \bibinfo {author}
  {\bibfnamefont {K.}~\bibnamefont {Geerlings}}, \bibinfo {author}
  {\bibfnamefont {T.}~\bibnamefont {Brecht}}, \bibinfo {author} {\bibfnamefont
  {K.}~\bibnamefont {Sliwa}}, \bibinfo {author} {\bibfnamefont
  {B.}~\bibnamefont {Abdo}}, \bibinfo {author} {\bibfnamefont {L.}~\bibnamefont
  {Frunzio}}, \bibinfo {author} {\bibfnamefont {S.~M.}\ \bibnamefont {Girvin}},
  \emph {et~al.},\ }\bibfield  {title} {\bibinfo {title} {Quantum back-action
  of an individual variable-strength measurement},\ }\href
  {https://doi.org/https://doi.org/10.1126/science.1226897} {\bibfield
  {journal} {\bibinfo  {journal} {Science}\ }\textbf {\bibinfo {volume}
  {339}},\ \bibinfo {pages} {178} (\bibinfo {year} {2013})}\BibitemShut
  {NoStop}%
\bibitem [{\citenamefont {Abdo}\ \emph {et~al.}(2021)\citenamefont {Abdo},
  \citenamefont {Jinka}, \citenamefont {Bronn}, \citenamefont {Olivadese},\
  and\ \citenamefont {Brink}}]{abdo2021high}%
  \BibitemOpen
  \bibfield  {author} {\bibinfo {author} {\bibfnamefont {B.}~\bibnamefont
  {Abdo}}, \bibinfo {author} {\bibfnamefont {O.}~\bibnamefont {Jinka}},
  \bibinfo {author} {\bibfnamefont {N.~T.}\ \bibnamefont {Bronn}}, \bibinfo
  {author} {\bibfnamefont {S.}~\bibnamefont {Olivadese}},\ and\ \bibinfo
  {author} {\bibfnamefont {M.}~\bibnamefont {Brink}},\ }\bibfield  {title}
  {\bibinfo {title} {High-fidelity qubit readout using interferometric
  directional josephson devices},\ }\href
  {https://doi.org/https://doi.org/10.1103/PRXQuantum.2.040360} {\bibfield
  {journal} {\bibinfo  {journal} {PRX Quantum}\ }\textbf {\bibinfo {volume}
  {2}},\ \bibinfo {pages} {040360} (\bibinfo {year} {2021})}\BibitemShut
  {NoStop}%
\bibitem [{\citenamefont {Leib}\ \emph {et~al.}(2016)\citenamefont {Leib},
  \citenamefont {Zoller},\ and\ \citenamefont {Lechner}}]{leib2016transmon}%
  \BibitemOpen
  \bibfield  {author} {\bibinfo {author} {\bibfnamefont {M.}~\bibnamefont
  {Leib}}, \bibinfo {author} {\bibfnamefont {P.}~\bibnamefont {Zoller}},\ and\
  \bibinfo {author} {\bibfnamefont {W.}~\bibnamefont {Lechner}},\ }\bibfield
  {title} {\bibinfo {title} {A transmon quantum annealer: decomposing many-body
  ising constraints into pair interactions},\ }\href
  {https://doi.org/10.1088/2058-9565/1/1/015008} {\bibfield  {journal}
  {\bibinfo  {journal} {Quantum Science and Technology}\ }\textbf {\bibinfo
  {volume} {1}},\ \bibinfo {pages} {015008} (\bibinfo {year}
  {2016})}\BibitemShut {NoStop}%
\bibitem [{\citenamefont {Roy}\ \emph {et~al.}(2017)\citenamefont {Roy},
  \citenamefont {Kundu}, \citenamefont {Chand}, \citenamefont {Hazra},
  \citenamefont {Nehra}, \citenamefont {Cosmic}, \citenamefont {Ranadive},
  \citenamefont {Patankar}, \citenamefont {Damle},\ and\ \citenamefont
  {Vijay}}]{roy2017implementation}%
  \BibitemOpen
  \bibfield  {author} {\bibinfo {author} {\bibfnamefont {T.}~\bibnamefont
  {Roy}}, \bibinfo {author} {\bibfnamefont {S.}~\bibnamefont {Kundu}}, \bibinfo
  {author} {\bibfnamefont {M.}~\bibnamefont {Chand}}, \bibinfo {author}
  {\bibfnamefont {S.}~\bibnamefont {Hazra}}, \bibinfo {author} {\bibfnamefont
  {N.}~\bibnamefont {Nehra}}, \bibinfo {author} {\bibfnamefont
  {R.}~\bibnamefont {Cosmic}}, \bibinfo {author} {\bibfnamefont
  {A.}~\bibnamefont {Ranadive}}, \bibinfo {author} {\bibfnamefont {M.~P.}\
  \bibnamefont {Patankar}}, \bibinfo {author} {\bibfnamefont {K.}~\bibnamefont
  {Damle}},\ and\ \bibinfo {author} {\bibfnamefont {R.}~\bibnamefont {Vijay}},\
  }\bibfield  {title} {\bibinfo {title} {Implementation of pairwise
  longitudinal coupling in a three-qubit superconducting circuit},\ }\href
  {https://doi.org/https://doi.org/10.1103/PhysRevApplied.7.054025} {\bibfield
  {journal} {\bibinfo  {journal} {Physical Review Applied}\ }\textbf {\bibinfo
  {volume} {7}},\ \bibinfo {pages} {054025} (\bibinfo {year}
  {2017})}\BibitemShut {NoStop}%
\bibitem [{\citenamefont {Liu}\ \emph {et~al.}(2017)\citenamefont {Liu},
  \citenamefont {Chien}, \citenamefont {Cao}, \citenamefont {Lanes},
  \citenamefont {Alpern}, \citenamefont {Pekker},\ and\ \citenamefont
  {Hatridge}}]{liu2017josephson}%
  \BibitemOpen
  \bibfield  {author} {\bibinfo {author} {\bibfnamefont {G.}~\bibnamefont
  {Liu}}, \bibinfo {author} {\bibfnamefont {T.-C.}\ \bibnamefont {Chien}},
  \bibinfo {author} {\bibfnamefont {X.}~\bibnamefont {Cao}}, \bibinfo {author}
  {\bibfnamefont {O.}~\bibnamefont {Lanes}}, \bibinfo {author} {\bibfnamefont
  {E.}~\bibnamefont {Alpern}}, \bibinfo {author} {\bibfnamefont
  {D.}~\bibnamefont {Pekker}},\ and\ \bibinfo {author} {\bibfnamefont
  {M.}~\bibnamefont {Hatridge}},\ }\bibfield  {title} {\bibinfo {title}
  {Josephson parametric converter saturation and higher order effects},\
  }\bibfield  {journal} {\bibinfo  {journal} {Applied Physics Letters}\
  }\textbf {\bibinfo {volume} {111}},\ \href
  {https://doi.org/10.1063/1.5003032} {10.1063/1.5003032} (\bibinfo {year}
  {2017})\BibitemShut {NoStop}%
\bibitem [{\citenamefont {Kim}\ \emph {et~al.}(2023)\citenamefont {Kim},
  \citenamefont {Lee}, \citenamefont {Kim}, \citenamefont {Jeong},\ and\
  \citenamefont {Kim}}]{kim2023squeezing}%
  \BibitemOpen
  \bibfield  {author} {\bibinfo {author} {\bibfnamefont {D.~H.}\ \bibnamefont
  {Kim}}, \bibinfo {author} {\bibfnamefont {S.-Y.}\ \bibnamefont {Lee}},
  \bibinfo {author} {\bibfnamefont {Z.}~\bibnamefont {Kim}}, \bibinfo {author}
  {\bibfnamefont {T.}~\bibnamefont {Jeong}},\ and\ \bibinfo {author}
  {\bibfnamefont {D.~Y.}\ \bibnamefont {Kim}},\ }\bibfield  {title} {\bibinfo
  {title} {Squeezing limit of the josephson ring modulator as a nondegenerate
  parametric amplifier},\ }\href
  {https://doi.org/https://doi.org/10.1103/PhysRevApplied.19.034015} {\bibfield
   {journal} {\bibinfo  {journal} {Physical Review Applied}\ }\textbf {\bibinfo
  {volume} {19}},\ \bibinfo {pages} {034015} (\bibinfo {year}
  {2023})}\BibitemShut {NoStop}%
\bibitem [{\citenamefont {Parker}\ \emph {et~al.}(2022)\citenamefont {Parker},
  \citenamefont {Savytskyi}, \citenamefont {Vine}, \citenamefont {Laucht},
  \citenamefont {Duty}, \citenamefont {Morello}, \citenamefont {Grimsmo},\ and\
  \citenamefont {Pla}}]{parker2022degenerate}%
  \BibitemOpen
  \bibfield  {author} {\bibinfo {author} {\bibfnamefont {D.~J.}\ \bibnamefont
  {Parker}}, \bibinfo {author} {\bibfnamefont {M.}~\bibnamefont {Savytskyi}},
  \bibinfo {author} {\bibfnamefont {W.}~\bibnamefont {Vine}}, \bibinfo {author}
  {\bibfnamefont {A.}~\bibnamefont {Laucht}}, \bibinfo {author} {\bibfnamefont
  {T.}~\bibnamefont {Duty}}, \bibinfo {author} {\bibfnamefont {A.}~\bibnamefont
  {Morello}}, \bibinfo {author} {\bibfnamefont {A.~L.}\ \bibnamefont
  {Grimsmo}},\ and\ \bibinfo {author} {\bibfnamefont {J.~J.}\ \bibnamefont
  {Pla}},\ }\bibfield  {title} {\bibinfo {title} {Degenerate parametric
  amplification via three-wave mixing using kinetic inductance},\ }\href
  {https://doi.org/https://doi.org/10.1103/PhysRevApplied.17.034064} {\bibfield
   {journal} {\bibinfo  {journal} {Physical Review Applied}\ }\textbf {\bibinfo
  {volume} {17}},\ \bibinfo {pages} {034064} (\bibinfo {year}
  {2022})}\BibitemShut {NoStop}%
\bibitem [{\citenamefont {Vaartjes}\ \emph {et~al.}(2023)\citenamefont
  {Vaartjes}, \citenamefont {Kringh{\o}j}, \citenamefont {Vine}, \citenamefont
  {Day}, \citenamefont {Morello},\ and\ \citenamefont
  {Pla}}]{vaartjes2023strong}%
  \BibitemOpen
  \bibfield  {author} {\bibinfo {author} {\bibfnamefont {A.}~\bibnamefont
  {Vaartjes}}, \bibinfo {author} {\bibfnamefont {A.}~\bibnamefont
  {Kringh{\o}j}}, \bibinfo {author} {\bibfnamefont {W.}~\bibnamefont {Vine}},
  \bibinfo {author} {\bibfnamefont {T.}~\bibnamefont {Day}}, \bibinfo {author}
  {\bibfnamefont {A.}~\bibnamefont {Morello}},\ and\ \bibinfo {author}
  {\bibfnamefont {J.~J.}\ \bibnamefont {Pla}},\ }\bibfield  {title} {\bibinfo
  {title} {Strong microwave squeezing above 1 tesla and 1 kelvin},\ }\bibfield
  {journal} {\bibinfo  {journal} {arXiv preprint arXiv:2311.07968}\ }\href
  {https://doi.org/10.48550/arXiv.2311.07968} {10.48550/arXiv.2311.07968}
  (\bibinfo {year} {2023})\BibitemShut {NoStop}%
\bibitem [{\citenamefont {Khalifa}\ and\ \citenamefont
  {Salfi}(2023)}]{khalifa2023nonlinearity}%
  \BibitemOpen
  \bibfield  {author} {\bibinfo {author} {\bibfnamefont {M.}~\bibnamefont
  {Khalifa}}\ and\ \bibinfo {author} {\bibfnamefont {J.}~\bibnamefont
  {Salfi}},\ }\bibfield  {title} {\bibinfo {title} {Nonlinearity and parametric
  amplification of superconducting nanowire resonators in magnetic field},\
  }\href {https://doi.org/https://doi.org/10.1103/PhysRevApplied.19.034024}
  {\bibfield  {journal} {\bibinfo  {journal} {Physical Review Applied}\
  }\textbf {\bibinfo {volume} {19}},\ \bibinfo {pages} {034024} (\bibinfo
  {year} {2023})}\BibitemShut {NoStop}%
\bibitem [{\citenamefont {Xu}\ \emph {et~al.}(2023)\citenamefont {Xu},
  \citenamefont {Cheng}, \citenamefont {Wu}, \citenamefont {Liu},\ and\
  \citenamefont {Tang}}]{xu2023magnetic}%
  \BibitemOpen
  \bibfield  {author} {\bibinfo {author} {\bibfnamefont {M.}~\bibnamefont
  {Xu}}, \bibinfo {author} {\bibfnamefont {R.}~\bibnamefont {Cheng}}, \bibinfo
  {author} {\bibfnamefont {Y.}~\bibnamefont {Wu}}, \bibinfo {author}
  {\bibfnamefont {G.}~\bibnamefont {Liu}},\ and\ \bibinfo {author}
  {\bibfnamefont {H.~X.}\ \bibnamefont {Tang}},\ }\bibfield  {title} {\bibinfo
  {title} {Magnetic field-resilient quantum-limited parametric amplifier},\
  }\href {https://doi.org/https://doi.org/10.1103/PRXQuantum.4.010322}
  {\bibfield  {journal} {\bibinfo  {journal} {PRX Quantum}\ }\textbf {\bibinfo
  {volume} {4}},\ \bibinfo {pages} {010322} (\bibinfo {year}
  {2023})}\BibitemShut {NoStop}%
\bibitem [{\citenamefont {Frasca}\ \emph {et~al.}(2024)\citenamefont {Frasca},
  \citenamefont {Roy}, \citenamefont {Beaulieu},\ and\ \citenamefont
  {Scarlino}}]{frasca2024three}%
  \BibitemOpen
  \bibfield  {author} {\bibinfo {author} {\bibfnamefont {S.}~\bibnamefont
  {Frasca}}, \bibinfo {author} {\bibfnamefont {C.}~\bibnamefont {Roy}},
  \bibinfo {author} {\bibfnamefont {G.}~\bibnamefont {Beaulieu}},\ and\
  \bibinfo {author} {\bibfnamefont {P.}~\bibnamefont {Scarlino}},\ }\bibfield
  {title} {\bibinfo {title} {Three-wave-mixing quantum-limited kinetic
  inductance parametric amplifier operating at 6 t near 1 k},\ }\href
  {https://doi.org/10.1103/PhysRevApplied.21.024011} {\bibfield  {journal}
  {\bibinfo  {journal} {Phys. Rev. Appl.}\ }\textbf {\bibinfo {volume} {21}},\
  \bibinfo {pages} {024011} (\bibinfo {year} {2024})}\BibitemShut {NoStop}%
\bibitem [{\citenamefont {Splitthoff}\ \emph {et~al.}(2024)\citenamefont
  {Splitthoff}, \citenamefont {Wesdorp}, \citenamefont {Pita-Vidal},
  \citenamefont {Bargerbos}, \citenamefont {Liu},\ and\ \citenamefont
  {Andersen}}]{splitthoff2024gate}%
  \BibitemOpen
  \bibfield  {author} {\bibinfo {author} {\bibfnamefont {L.~J.}\ \bibnamefont
  {Splitthoff}}, \bibinfo {author} {\bibfnamefont {J.~J.}\ \bibnamefont
  {Wesdorp}}, \bibinfo {author} {\bibfnamefont {M.}~\bibnamefont {Pita-Vidal}},
  \bibinfo {author} {\bibfnamefont {A.}~\bibnamefont {Bargerbos}}, \bibinfo
  {author} {\bibfnamefont {Y.}~\bibnamefont {Liu}},\ and\ \bibinfo {author}
  {\bibfnamefont {C.~K.}\ \bibnamefont {Andersen}},\ }\bibfield  {title}
  {\bibinfo {title} {Gate-tunable kinetic inductance parametric amplifier},\
  }\href {https://doi.org/10.1103/PhysRevApplied.21.014052} {\bibfield
  {journal} {\bibinfo  {journal} {Phys. Rev. Appl.}\ }\textbf {\bibinfo
  {volume} {21}},\ \bibinfo {pages} {014052} (\bibinfo {year}
  {2024})}\BibitemShut {NoStop}%
\bibitem [{\citenamefont {Malnou}\ \emph {et~al.}(2022)\citenamefont {Malnou},
  \citenamefont {Aumentado}, \citenamefont {Vissers}, \citenamefont {Wheeler},
  \citenamefont {Hubmayr}, \citenamefont {Ullom},\ and\ \citenamefont
  {Gao}}]{malnou2022performance}%
  \BibitemOpen
  \bibfield  {author} {\bibinfo {author} {\bibfnamefont {M.}~\bibnamefont
  {Malnou}}, \bibinfo {author} {\bibfnamefont {J.}~\bibnamefont {Aumentado}},
  \bibinfo {author} {\bibfnamefont {M.}~\bibnamefont {Vissers}}, \bibinfo
  {author} {\bibfnamefont {J.}~\bibnamefont {Wheeler}}, \bibinfo {author}
  {\bibfnamefont {J.}~\bibnamefont {Hubmayr}}, \bibinfo {author} {\bibfnamefont
  {J.}~\bibnamefont {Ullom}},\ and\ \bibinfo {author} {\bibfnamefont
  {J.}~\bibnamefont {Gao}},\ }\bibfield  {title} {\bibinfo {title} {Performance
  of a kinetic inductance traveling-wave parametric amplifier at 4 kelvin:
  Toward an alternative to semiconductor amplifiers},\ }\href
  {https://doi.org/https://doi.org/10.1103/PhysRevApplied.17.044009} {\bibfield
   {journal} {\bibinfo  {journal} {Physical Review Applied}\ }\textbf {\bibinfo
  {volume} {17}},\ \bibinfo {pages} {044009} (\bibinfo {year}
  {2022})}\BibitemShut {NoStop}%
\bibitem [{\citenamefont {Mohamed}\ \emph {et~al.}(2024)\citenamefont
  {Mohamed}, \citenamefont {Zohari}, \citenamefont {Pla}, \citenamefont
  {Barclay},\ and\ \citenamefont {Barzanjeh}}]{mohamed2023selective}%
  \BibitemOpen
  \bibfield  {author} {\bibinfo {author} {\bibfnamefont {A.}~\bibnamefont
  {Mohamed}}, \bibinfo {author} {\bibfnamefont {E.}~\bibnamefont {Zohari}},
  \bibinfo {author} {\bibfnamefont {J.~J.}\ \bibnamefont {Pla}}, \bibinfo
  {author} {\bibfnamefont {P.~E.}\ \bibnamefont {Barclay}},\ and\ \bibinfo
  {author} {\bibfnamefont {S.}~\bibnamefont {Barzanjeh}},\ }\bibfield  {title}
  {\bibinfo {title} {Selective single- and double-mode quantum-limited
  amplifier},\ }\href {https://doi.org/10.1103/PhysRevApplied.21.064052}
  {\bibfield  {journal} {\bibinfo  {journal} {Phys. Rev. Appl.}\ }\textbf
  {\bibinfo {volume} {21}},\ \bibinfo {pages} {064052} (\bibinfo {year}
  {2024})}\BibitemShut {NoStop}%
\bibitem [{\citenamefont {Wu}\ \emph {et~al.}(2024)\citenamefont {Wu},
  \citenamefont {Xu},\ and\ \citenamefont {Tang}}]{wu2023junction}%
  \BibitemOpen
  \bibfield  {author} {\bibinfo {author} {\bibfnamefont {Y.}~\bibnamefont
  {Wu}}, \bibinfo {author} {\bibfnamefont {M.}~\bibnamefont {Xu}},\ and\
  \bibinfo {author} {\bibfnamefont {H.~X.}\ \bibnamefont {Tang}},\ }\bibfield
  {title} {\bibinfo {title} {Junction-free microwave two-mode radiation from a
  kinetic inductance nanowire},\ }\href
  {https://doi.org/10.1103/PhysRevApplied.21.014029} {\bibfield  {journal}
  {\bibinfo  {journal} {Phys. Rev. Appl.}\ }\textbf {\bibinfo {volume} {21}},\
  \bibinfo {pages} {014029} (\bibinfo {year} {2024})}\BibitemShut {NoStop}%
\bibitem [{\citenamefont {Bergeal}\ \emph
  {et~al.}(2010{\natexlab{b}})\citenamefont {Bergeal}, \citenamefont
  {Schackert}, \citenamefont {Metcalfe}, \citenamefont {Vijay}, \citenamefont
  {Manucharyan}, \citenamefont {Frunzio}, \citenamefont {Prober}, \citenamefont
  {Schoelkopf}, \citenamefont {Girvin},\ and\ \citenamefont
  {Devoret}}]{bergeal2010phase}%
  \BibitemOpen
  \bibfield  {author} {\bibinfo {author} {\bibfnamefont {N.}~\bibnamefont
  {Bergeal}}, \bibinfo {author} {\bibfnamefont {F.}~\bibnamefont {Schackert}},
  \bibinfo {author} {\bibfnamefont {M.}~\bibnamefont {Metcalfe}}, \bibinfo
  {author} {\bibfnamefont {R.}~\bibnamefont {Vijay}}, \bibinfo {author}
  {\bibfnamefont {V.}~\bibnamefont {Manucharyan}}, \bibinfo {author}
  {\bibfnamefont {L.}~\bibnamefont {Frunzio}}, \bibinfo {author} {\bibfnamefont
  {D.}~\bibnamefont {Prober}}, \bibinfo {author} {\bibfnamefont
  {R.}~\bibnamefont {Schoelkopf}}, \bibinfo {author} {\bibfnamefont
  {S.}~\bibnamefont {Girvin}},\ and\ \bibinfo {author} {\bibfnamefont
  {M.}~\bibnamefont {Devoret}},\ }\bibfield  {title} {\bibinfo {title}
  {Phase-preserving amplification near the quantum limit with a josephson ring
  modulator},\ }\href {https://doi.org/https://doi.org/10.1038/nature09035}
  {\bibfield  {journal} {\bibinfo  {journal} {Nature}\ }\textbf {\bibinfo
  {volume} {465}},\ \bibinfo {pages} {64} (\bibinfo {year}
  {2010}{\natexlab{b}})}\BibitemShut {NoStop}%
\bibitem [{\citenamefont {Eom}\ \emph {et~al.}(2012)\citenamefont {Eom},
  \citenamefont {Day}, \citenamefont {LeDuc},\ and\ \citenamefont
  {Zmuidzinas}}]{eom2012wideband}%
  \BibitemOpen
  \bibfield  {author} {\bibinfo {author} {\bibfnamefont {B.~H.}\ \bibnamefont
  {Eom}}, \bibinfo {author} {\bibfnamefont {P.~K.}\ \bibnamefont {Day}},
  \bibinfo {author} {\bibfnamefont {H.~G.}\ \bibnamefont {LeDuc}},\ and\
  \bibinfo {author} {\bibfnamefont {J.}~\bibnamefont {Zmuidzinas}},\ }\bibfield
   {title} {\bibinfo {title} {A wideband, low-noise superconducting amplifier
  with high dynamic range},\ }\href
  {https://doi.org/https://doi.org/10.1038/nphys2356} {\bibfield  {journal}
  {\bibinfo  {journal} {Nature Physics}\ }\textbf {\bibinfo {volume} {8}},\
  \bibinfo {pages} {623} (\bibinfo {year} {2012})}\BibitemShut {NoStop}%
\bibitem [{\citenamefont {G{\"o}ppl}\ \emph {et~al.}(2008)\citenamefont
  {G{\"o}ppl}, \citenamefont {Fragner}, \citenamefont {Baur}, \citenamefont
  {Bianchetti}, \citenamefont {Filipp}, \citenamefont {Fink}, \citenamefont
  {Leek}, \citenamefont {Puebla}, \citenamefont {Steffen},\ and\ \citenamefont
  {Wallraff}}]{goppl2008coplanar}%
  \BibitemOpen
  \bibfield  {author} {\bibinfo {author} {\bibfnamefont {M.}~\bibnamefont
  {G{\"o}ppl}}, \bibinfo {author} {\bibfnamefont {A.}~\bibnamefont {Fragner}},
  \bibinfo {author} {\bibfnamefont {M.}~\bibnamefont {Baur}}, \bibinfo {author}
  {\bibfnamefont {R.}~\bibnamefont {Bianchetti}}, \bibinfo {author}
  {\bibfnamefont {S.}~\bibnamefont {Filipp}}, \bibinfo {author} {\bibfnamefont
  {J.~M.}\ \bibnamefont {Fink}}, \bibinfo {author} {\bibfnamefont {P.~J.}\
  \bibnamefont {Leek}}, \bibinfo {author} {\bibfnamefont {G.}~\bibnamefont
  {Puebla}}, \bibinfo {author} {\bibfnamefont {L.}~\bibnamefont {Steffen}},\
  and\ \bibinfo {author} {\bibfnamefont {A.}~\bibnamefont {Wallraff}},\
  }\bibfield  {title} {\bibinfo {title} {Coplanar waveguide resonators for
  circuit quantum electrodynamics},\ }\bibfield  {journal} {\bibinfo  {journal}
  {Journal of Applied Physics}\ }\textbf {\bibinfo {volume} {104}},\ \href
  {https://doi.org/10.1063/1.3010859} {10.1063/1.3010859} (\bibinfo {year}
  {2008})\BibitemShut {NoStop}%
\bibitem [{\citenamefont {Samkharadze}\ \emph {et~al.}(2016)\citenamefont
  {Samkharadze}, \citenamefont {Bruno}, \citenamefont {Scarlino}, \citenamefont
  {Zheng}, \citenamefont {DiVincenzo}, \citenamefont {DiCarlo},\ and\
  \citenamefont {Vandersypen}}]{samkharadze2016high}%
  \BibitemOpen
  \bibfield  {author} {\bibinfo {author} {\bibfnamefont {N.}~\bibnamefont
  {Samkharadze}}, \bibinfo {author} {\bibfnamefont {A.}~\bibnamefont {Bruno}},
  \bibinfo {author} {\bibfnamefont {P.}~\bibnamefont {Scarlino}}, \bibinfo
  {author} {\bibfnamefont {G.}~\bibnamefont {Zheng}}, \bibinfo {author}
  {\bibfnamefont {D.}~\bibnamefont {DiVincenzo}}, \bibinfo {author}
  {\bibfnamefont {L.}~\bibnamefont {DiCarlo}},\ and\ \bibinfo {author}
  {\bibfnamefont {L.}~\bibnamefont {Vandersypen}},\ }\bibfield  {title}
  {\bibinfo {title} {High-kinetic-inductance superconducting nanowire
  resonators for circuit qed in a magnetic field},\ }\href
  {https://doi.org/10.1103/PhysRevApplied.5.044004} {\bibfield  {journal}
  {\bibinfo  {journal} {Physical Review Applied}\ }\textbf {\bibinfo {volume}
  {5}},\ \bibinfo {pages} {044004} (\bibinfo {year} {2016})}\BibitemShut
  {NoStop}%
\bibitem [{\citenamefont {Abdo}\ \emph {et~al.}(2011)\citenamefont {Abdo},
  \citenamefont {Schackert}, \citenamefont {Hatridge}, \citenamefont
  {Rigetti},\ and\ \citenamefont {Devoret}}]{abdo2011josephson}%
  \BibitemOpen
  \bibfield  {author} {\bibinfo {author} {\bibfnamefont {B.}~\bibnamefont
  {Abdo}}, \bibinfo {author} {\bibfnamefont {F.}~\bibnamefont {Schackert}},
  \bibinfo {author} {\bibfnamefont {M.}~\bibnamefont {Hatridge}}, \bibinfo
  {author} {\bibfnamefont {C.}~\bibnamefont {Rigetti}},\ and\ \bibinfo {author}
  {\bibfnamefont {M.}~\bibnamefont {Devoret}},\ }\bibfield  {title} {\bibinfo
  {title} {{Josephson amplifier for qubit readout}},\ }\href
  {https://doi.org/10.1063/1.3653473} {\bibfield  {journal} {\bibinfo
  {journal} {Applied Physics Letters}\ }\textbf {\bibinfo {volume} {99}},\
  \bibinfo {pages} {162506} (\bibinfo {year} {2011})}\BibitemShut {NoStop}%
\bibitem [{\citenamefont {Roch}\ \emph {et~al.}(2012)\citenamefont {Roch},
  \citenamefont {Flurin}, \citenamefont {Nguyen}, \citenamefont {Morfin},
  \citenamefont {Campagne-Ibarcq}, \citenamefont {Devoret},\ and\ \citenamefont
  {Huard}}]{roch2012widely}%
  \BibitemOpen
  \bibfield  {author} {\bibinfo {author} {\bibfnamefont {N.}~\bibnamefont
  {Roch}}, \bibinfo {author} {\bibfnamefont {E.}~\bibnamefont {Flurin}},
  \bibinfo {author} {\bibfnamefont {F.}~\bibnamefont {Nguyen}}, \bibinfo
  {author} {\bibfnamefont {P.}~\bibnamefont {Morfin}}, \bibinfo {author}
  {\bibfnamefont {P.}~\bibnamefont {Campagne-Ibarcq}}, \bibinfo {author}
  {\bibfnamefont {M.~H.}\ \bibnamefont {Devoret}},\ and\ \bibinfo {author}
  {\bibfnamefont {B.}~\bibnamefont {Huard}},\ }\bibfield  {title} {\bibinfo
  {title} {Widely tunable, nondegenerate three-wave mixing microwave device
  operating near the quantum limit},\ }\href
  {https://doi.org/10.1103/PhysRevLett.108.147701} {\bibfield  {journal}
  {\bibinfo  {journal} {Phys. Rev. Lett.}\ }\textbf {\bibinfo {volume} {108}},\
  \bibinfo {pages} {147701} (\bibinfo {year} {2012})}\BibitemShut {NoStop}%
\bibitem [{\citenamefont {Schackert}\ \emph {et~al.}(2013)\citenamefont
  {Schackert}, \citenamefont {Roy}, \citenamefont {Hatridge}, \citenamefont
  {Devoret},\ and\ \citenamefont {Stone}}]{schackert2013three}%
  \BibitemOpen
  \bibfield  {author} {\bibinfo {author} {\bibfnamefont {F.}~\bibnamefont
  {Schackert}}, \bibinfo {author} {\bibfnamefont {A.}~\bibnamefont {Roy}},
  \bibinfo {author} {\bibfnamefont {M.}~\bibnamefont {Hatridge}}, \bibinfo
  {author} {\bibfnamefont {M.~H.}\ \bibnamefont {Devoret}},\ and\ \bibinfo
  {author} {\bibfnamefont {A.~D.}\ \bibnamefont {Stone}},\ }\bibfield  {title}
  {\bibinfo {title} {Three-wave mixing with three incoming waves: signal-idler
  coherent attenuation and gain enhancement in a parametric amplifier},\ }\href
  {https://doi.org/https://doi.org/10.1103/PhysRevLett.111.073903} {\bibfield
  {journal} {\bibinfo  {journal} {Physical review letters}\ }\textbf {\bibinfo
  {volume} {111}},\ \bibinfo {pages} {073903} (\bibinfo {year}
  {2013})}\BibitemShut {NoStop}%
\bibitem [{\citenamefont {Clem}\ and\ \citenamefont
  {Kogan}(2012)}]{clem2012kinetic}%
  \BibitemOpen
  \bibfield  {author} {\bibinfo {author} {\bibfnamefont {J.~R.}\ \bibnamefont
  {Clem}}\ and\ \bibinfo {author} {\bibfnamefont {V.}~\bibnamefont {Kogan}},\
  }\bibfield  {title} {\bibinfo {title} {Kinetic impedance and depairing in
  thin and narrow superconducting films},\ }\href
  {https://doi.org/https://doi.org/10.1103/PhysRevB.86.174521} {\bibfield
  {journal} {\bibinfo  {journal} {Physical Review B}\ }\textbf {\bibinfo
  {volume} {86}},\ \bibinfo {pages} {174521} (\bibinfo {year}
  {2012})}\BibitemShut {NoStop}%
\bibitem [{\citenamefont {Gardiner}\ and\ \citenamefont
  {Collett}(1985)}]{GardinerCollett1985}%
  \BibitemOpen
  \bibfield  {author} {\bibinfo {author} {\bibfnamefont {C.~W.}\ \bibnamefont
  {Gardiner}}\ and\ \bibinfo {author} {\bibfnamefont {M.~J.}\ \bibnamefont
  {Collett}},\ }\bibfield  {title} {\bibinfo {title} {Input and output in
  damped quantum systems: Quantum stochastic differential equations and the
  master equation},\ }\href {https://doi.org/10.1103/PhysRevA.31.3761}
  {\bibfield  {journal} {\bibinfo  {journal} {Phys. Rev. A}\ }\textbf {\bibinfo
  {volume} {31}},\ \bibinfo {pages} {3761} (\bibinfo {year}
  {1985})}\BibitemShut {NoStop}%
\end{thebibliography}
%

\end{document}